\documentclass[aps,prb,twocolumn,superscriptaddress,floatfix,letter]{revtex4-1}
\usepackage{epsfig,amsmath,amssymb,color,dsfont,physics}
\definecolor{jadclr}{rgb}{0,0.5,0}
\definecolor{jadcolor}{rgb}{0.5,0,0}
\definecolor{philipp}{rgb}{1,.4,.3}

\usepackage[bookmarks=true,colorlinks,linkcolor=jadclr,urlcolor=jadcolor,citecolor=jadcolor]{hyperref}
\usepackage{bm}
\usepackage{booktabs}
\usepackage{enumerate}
\newcommand{\RR}{\mathbb{R}}
\newcommand{\CC}{\mathbb{C}}

\usepackage{array}
\usepackage{dsfont}
\usepackage{microtype}
\usepackage[normalem]{ulem}

\usepackage{mathtools}

\DeclarePairedDelimiter{\floor}{\lfloor}{\rfloor}
\usepackage{mathrsfs}
\makeatletter
\newcommand{\thickhline}{
    \noalign {\ifnum 0=`}\fi \hrule height 1pt
    \futurelet \reserved@a \@xhline
}
\newcolumntype{"}{@{\hskip\tabcolsep\vrule width 1pt\hskip\tabcolsep}}
\makeatother
\begin{document}

\title{Out-of-equilibrium phase diagram of long-range superconductors}
\author{Philipp Uhrich}
\affiliation{Kirchhoff Institute for Physics, Heidelberg University, 69120 Heidelberg, Germany}
\affiliation{Institute for Theoretical Physics, Heidelberg University, 69120 Heidelberg, Germany}
\affiliation{Institute of Physics, \'Ecole Polytechnique F\'ed\'erale de Lausanne, Lausanne 1015, Switzerland}

\author{Nicol\`o Defenu}
\affiliation{Institute for Theoretical Physics, Heidelberg University, 69120 Heidelberg, Germany}

\author{Rouhollah Jafari}
\affiliation{Department of Physics, Institute for Advanced Studies in Basic Sciences (IASBS), Zanjan 45137-66731, Iran}
\affiliation{Department of Physics, University of Gothenburg, SE 412 96 Gothenburg, Sweden}

\author{Jad C.~Halimeh}
\affiliation{Kirchhoff Institute for Physics, Heidelberg University, 69120 Heidelberg, Germany}
\affiliation{Institute for Theoretical Physics, Heidelberg University, 69120 Heidelberg, Germany}
\affiliation{Max Planck Institute for the Physics of Complex Systems, 01187 Dresden, Germany}

\date{\today}

\begin{abstract}
Within the ultimate goal of classifying universality in quantum many-body dynamics, understanding the relation between out-of-equilibrium and equilibrium criticality is a crucial objective. Models with power-law interactions exhibit rich well-understood critical behavior in equilibrium, but the out-of-equilibrium picture has remained incomplete, despite recent experimental progress. We construct the rich dynamical phase diagram of free-fermionic chains with power-law hopping and pairing, and provide analytic and numerical evidence showing a direct connection between nonanalyticities of the return rate and zero crossings of the string order parameter. Our results may explain the experimental observation of so-called \textit{accidental} dynamical vortices, which appear for quenches within the same topological phase of the Haldane model, as reported in [Fl\"aschner \textit{et al.}, Nature Physics \textbf{14}, 265 (2018)]. Our work is readily applicable to modern ultracold-atom experiments, not least because state-of-the-art quantum gas microscopes can now reliably measure the string order parameter, which, as we show, can serve as an indicator of dynamical criticality.
\end{abstract}

\maketitle

\section{Introduction}
Criticality and universality in equilibrium\cite{Cardy_book,Sachdev_book,Ma_book} are well-established concepts thanks in large part to the renormalization group.\cite{Wilson1971a,Wilson1971b} Universality classes in out-of-equilibrium physics have also been developed for classical systems.\cite{Hohenberg1977,Taeuber_book} Nevertheless, criticality and universality in out-of-equilibrium quantum many-body systems are still much less understood concepts that are currently at the forefront of research efforts in condensed matter and cold-atom physics. 

Among the prominent thrusts of this research effort lies the phenomenon of dynamical phase transitions,\cite{Zvyagin2016,Mori2018} which fall into two major categories when considering sudden quenches of a quantum many-body system. The first is characterized by a local order parameter, whose long-time behavior determines the dynamical phase of the steady state.\cite{Moeckel2008,Eckstein2008,Eckstein2009,Sciolla2010,Sciolla2011,Gambassi2011,Sciolla2013,Chandran2013,Halimeh2017,Halimeh2017b,Zauner2017,Homrighausen2017,Lang2017,Lang2018,Halimeh2018a,Hashizume2018,Mori2019} In principle, this type of dynamical phase transition separates a ferromagnetic from a paramagnetic steady state in the wake of a quench, and this has been recently observed experimentally in trapped-ion setups.\cite{Zhang2017} However, in cases where the system has no finite-temperature phase transition and, therefore, always ends up in a paramagnetic steady state in the infinite-time limit, this dynamical phase transition can still be defined based on the decay behavior of the order parameter.\cite{Altman2002,Barmettler2009,Heyl2014,Halimeh2018a,Hashizume2018} For small enough quenches starting in an ordered state, the order parameter would decay asymptotically to zero without ever crossing zero. On the other hand, when the quench is large enough, the order parameter exhibits zero crossings as a function of time while forming an envelope that decays to zero in the long-time limit. Very interestingly, however, even when the model has no finite-temperature phase transition, the order parameter can exhibit persistent order at moderate-to-long times in the case of sufficiently small quenches in the presence of long-range interactions,\cite{Halimeh2017b} which has been shown to be due to domain-wall binding and confined dynamics.\cite{Liu2019,Halimeh2018a}

The second type of dynamical phase transition is based on nonanalyticities in the so-called Loschmidt return rate\cite{Heyl2013} (see Sec.~\ref{sec:dqpt} for an overview), which can be construed as a dynamical analog of the thermal free energy with complexified evolution time standing for inverse temperature. Consequently, nonanalyticities in the return rate are construed as \textit{dynamical quantum phase transitions} (DQPT) that occur at \textit{critical evolution times} after a quench. In the few years since its introduction,\cite{Heyl2013} DQPT has witnessed a flurry of research activity from the perspective of both theory\cite{Heyl_review,Mori2018} and experiment.\cite{Flaeschner2018,Jurcevic2017} Even though the seminal work of Ref.~\onlinecite{Heyl2013} showed in the case of the nearest-neighbor transverse-field Ising chain (TFIC) that nonanalyticities in the return rate occur only upon quenching across the quantum equilibrium critical point, soon thereafter it was indicated that this is neither a necessary nor a sufficient condition in other integrable and nonintegrable models.\cite{Andraschko2014,Vajna2014} Indeed, it was then firmly established that the dynamical critical point, which separates different phases of DQPT, is in general distinct from the quantum equilibrium critical point and is strongly dependent on the initial condition.\cite{Halimeh2017,Homrighausen2017,Lang2017} Additionally, this dynamical critical point was shown to coincide with that of the type of dynamical phase transition based on a local order parameter.\cite{Halimeh2017,Zauner2017,Homrighausen2017,Lang2017,Lang2018,Halimeh2018a,Hashizume2018} Also recently, the theory of DQPT has been extended to Floquet systems\cite{Kosior2018a,Kosior2018b,Yang2019} and models with many-body localized phases.\cite{Gurarie2018,Halimeh2019b} DQPTs in nonintegrable models have been studied in, for instance, Ref.~\onlinecite{Karrasch13}.

Free-fermionic models have been a fundamental ingredient to the comprehension of quantum critical behavior both in and out of equilibrium. The introduction of long-range pairing couplings deeply influences the equilibrium phase diagram,\cite{Vodola2014,Lepori2017,Viyuela2015,Viyuela18,Cats18} leading to peculiar singularities in the spectrum of the system\,\cite{Lepori2017And} and violations of conformal symmetries.\cite{Lepori2015,Lepori2016Gori,Lepori2017} Similarly, long-range hopping terms lead to the appearance of multiple Majorana edge modes.\cite{Alecce2017} Furthermore, while in the nearest-neighbor case the quantum critical behaviour of free-fermionic models maps onto that of quantum spins,\cite{Fradkin2013} the introduction of long-range interactions in the latter spoils such equivalence.\cite{Defenu2017,Defenu2016,Defenu2019wlr}  In the dynamical realm, long-range couplings induce a wide variety of peculiar features such as modified Lieb-Robinson bounds\,\cite{Hauke2013,Foss-Feig2015} and anomalous Kibble-Zurek scaling.\cite{Defenu2019univ} Finally, long-range (interacting) fermion models have also been studied in the context of phase coherence in $1$D superconductors.~\cite{Lobos13}

In the framework of DQPTs free-fermionic models have been at the center of theoretical and even experimental investigations. Indeed, the connection of DQPT to the underlying topological phase transition\cite{Budich2016} was first illustrated in Bogoliubov-de Gennes Hamiltonians, and one of the first experiments on DQPT implemented the Haldane model on a hexagonal lattice.\cite{Flaeschner2018} The relation of topology changing quenches and DQPTs was further studied for one- and two-dimensional systems in, for instance, Refs.~\onlinecite{Bhattacharya17a,Bhattacharya17b,Vajna15}. On the other hand, long-range interactions have proven to give rise to rich dynamical critical behavior,\cite{Halimeh2017} just as they do for equilibrium criticality.\cite{Dutta2001,Maghrebi2016} In particular, the DQPT picture qualitatively changes when long-range interactions are introduced. Domain-wall binding and confined dynamics at sufficiently small transverse-field strengths in long-range quantum Ising chains\cite{Halimeh2018a,Liu2019} lead to the appearance of anomalous cusps\cite{Halimeh2017,Zauner2017} in the return rate (see Sec.~\ref{sec:dqpt}), which would be entirely smooth otherwise. However, long-range quantum Ising chains are nonintegrable, and most DQPT studies on them have required advanced numerical treatment, such as the time-dependent variational principle (TDVP) within uniform matrix product states.\cite{Haegeman2011,Haegeman2016,Zauner2018} Consequently, the interest in free-fermionic models with long-range hopping or pairing has grown in recent years, as this allows the study of DQPT in an exactly solvable setting. Indeed, Dutta and Dutta\cite{Dutta2017} have recently investigated DQPT in the Kitaev chain with nearest-neighbor hopping and long-range pairing, while Defenu \textit{et al.}\cite{Defenu2019} have employed a truncated Jordan-Wigner transformation on the so-called extended long-range Ising chain to map it onto a free-fermionic model of long-range hopping and pairing both scaling as $1/r^\alpha$, with $r$ the inter-spin distance and $\alpha>1$. In this paper, we investigate a Kitaev chain with generalized power-law hopping and pairing profiles and draw a connection between the different dynamical phases of DQPT and the dynamic behavior of the string order parameter (SOP). String order parameters are related to hidden symmetry breaking and distinguish between topologically ordered and trivial equilibrium phases.\cite{Pollmann12} Comparing the dynamics of the SOP after a quench to that of the return rate can thus provide a natural link between DQPT and the underlying equilibrium topological phase transitions.

The remainder of the paper is organized as follows: In Sec.~\ref{sec:model}, we introduce our model and discuss its equilibrium phase diagram. In Sec.~\ref{sec:dqpt}, we provide a brief overview of the theory of DQPT. An analytic treatment with which we construct the dynamical phase diagram of the Kitaev chain with general power-law hopping and pairing terms is then presented in Sec.~\ref{sec:DPD}. In Sec.~\ref{sec:sop}, we derive the string order parameter for our model. Our results for the Loschmidt return rates and string order parameter dynamics are then discussed in Sec.~\ref{sec:results}. We conclude in Sec.~\ref{sec:conc}.

\section{Model}\label{sec:model}
The model we consider is the generalized Kac-normalized\cite{Kac1963} long-range Kitaev chain (LRKC) with power-law hopping and pairing profiles and closed boundary conditions. See Ref.~\onlinecite{Maity19} for a review. It is described by the Hamiltonian

\begin{align}\nonumber\label{eq:lrkc}
H=&\,-\sum_{j=1}^N\sum_{r=1}^{\frac{N}{2}-1}\big(J_rc_j^\dagger c_{j+r}+\Delta_rc_j^\dagger c_{j+r}^\dagger+\text{H.c.}\big)\\
&\,-h\sum_{j=1}^N\big(1-2c_j^\dagger c_j\big),
\end{align}
where
\begin{align}\label{e:pairhop}
&J_r=\frac{r^{-\alpha}}{\mathcal{N}_\alpha},\,\,\,\,\,\,\,\Delta_r=\frac{r^{-\beta}}{\mathcal{N}_\beta},
\end{align}
are the hopping and pairing profiles, respectively, with Kac normalization factors
\begin{align}
&\mathcal{N}_\alpha=\sum_{r=1}^{N/2-1}r^{-\alpha},\,\,\,\,\,\,\, \mathcal{N}_\beta=\sum_{r=1}^{N/2-1}r^{-\beta}.
\end{align}
Here and throughout $N$ denotes the system size, $\alpha,\beta>1$ the power-law exponents, $h$ the chemical-potential strength, and $c_j,c_j^\dagger$ the fermionic annihilation and creation operators, respectively, on site $j$ that satisfy the canonical anticommutation relations $\{c_l,c_j^\dagger\}=\delta_{l,j}$ and $\{c_l,c_j\}=0$. In~\eqref{e:pairhop} we have adopted the ring convention of Ref.~\onlinecite{Alecce2017}, which translates into $r$ ranging from $1$ up to only $N/2-1$ rather than to $N$ in~\eqref{eq:lrkc}. This is done in order to adequately deal with closed boundary conditions and effectively utilize the Fourier transformation
\begin{align}
c_j=\frac{1}{\sqrt{N}}\sum_k^\text{B.z.}c_k\mathrm{e}^{\mathrm{i}kj},
\end{align} 
into momentum space motivated by the translation invariance of~\eqref{eq:lrkc}. In the limit $N \to \infty$, the choice between periodic or anti-periodic boundaries is arbitrary as they yield the same momentum-space Hamiltonian

\begin{align}\nonumber
H=&\,\sum_k^\text{B.z.}\,\big[(c_k^\dagger c_k-c_{-k}c_{-k}^\dagger)(h-J_k)\\\label{eq:lrkc_momentum}
&\,+(c_k^\dagger c_{-k}^\dagger-c_{k}c_{-k})\Delta_k\big],
\end{align}
where the Fourier transforms of $J_r$ and $\Delta_r$ are

\begin{align}\label{eq:Jk}
J_k=\frac{1}{\zeta(\alpha)}\sum_{r=1}^\infty\frac{\cos(kr)}{r^\alpha}=\frac{\mathrm{Re[\,Li}_\alpha\left(e^{ik}\right)]}{2\zeta(\alpha)},\\\label{eq:Deltak}
\Delta_k=\frac{1}{\zeta(\beta)}\sum_{r=1}^\infty\frac{\sin(kr)}{r^\beta}=\frac{\mathrm{Im[\,Li}_\beta\left(e^{ik}\right)]}{2\zeta(\beta)},
\end{align}
respectively, and $\mathrm{Li}_s(z)$ denotes the polylogarithm of order $s$ and argument $z$.
We recall here that we are working in the thermodynamic limit $N\to\infty$, hence the Riemann Zeta functions $\zeta$ arise in these coefficients. The Riemann Zeta functions perform a Kac normalization that fixes the positive quantum equilibrium critical point of the model to $h_c^e=1$.

To solve for the spectrum of \eqref{eq:lrkc_momentum} one employs the Bogoliubov transformation

\begin{align}\label{eq:BogoTrafo}
c_k=\mathrm{i}\sin\frac{\theta_k}{2}\gamma_k+\cos\frac{\theta_k}{2}\gamma_{-k}^\dagger,
\end{align}
where $\gamma_k,\gamma_k^\dagger$ are the fermionic Bogoliubov annihilation and creation operators, respectively, with momentum $k$ obeying the canonical anticommutation relations $\{\gamma_k,\gamma_p^\dagger\}=\delta_{k,p}$ and $\{\gamma_k,\gamma_p\}=0$, and

\begin{align}
\theta_k=\arctan\frac{\Delta_k}{h-J_k}.
\end{align}
The transformation~\eqref{eq:BogoTrafo} diagonalizes~\eqref{eq:lrkc_momentum} into the form

\begin{align}\label{e:Ham}
H=&\,\sum_k^\text{B.z.}\omega_k(\gamma_k^\dagger \gamma_k-\gamma_{-k}\gamma_{-k}^\dagger),\\\label{e:Ham2}
\omega_k=&\,\sqrt{(h-J_k)^2+\Delta_k^2}.
\end{align}

\begin{figure}[!ht]
	\centering
	\includegraphics[width=\linewidth]{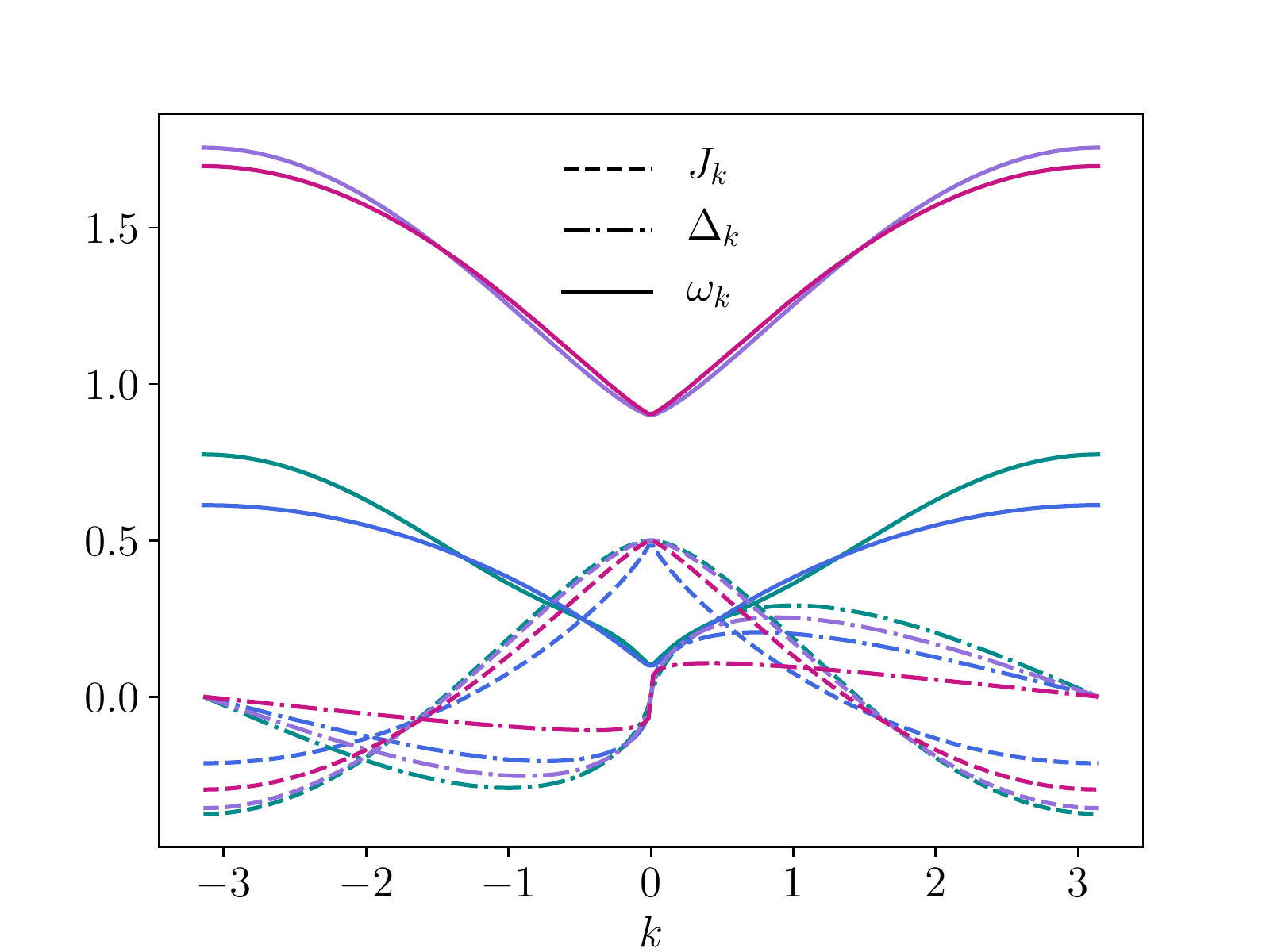}
	\caption{\label{f:dispersion}
	(Color online) Dispersion relation $\omega_k$ \eqref{e:Ham} together with the Fourier transforms of the hopping amplitude $J_k$ \eqref{eq:Jk} and pairing  amplitude $\Delta_k$ \eqref{eq:Deltak} for quasi-momenta $k$ within the first Brillouin zone. The dispersion relation is calculated with the following long-range hopping and pairing exponents $\alpha$ and $\beta$ and chemical potential $h$, listed from top to bottom of the left end of each solid curve: $(\alpha,\beta,h)=(2.8,1.7,1.4)$ in purple, $(\alpha,\beta,h)=(2.3,1.2,1.4)$ in violet, $(\alpha,\beta,h)=(3,1.9,0.4)$ in green, $(\alpha,\beta,h)=(1.8,1.5,0.4)$ in blue.
}
\end{figure}

\section{Dynamical quantum phase transitions}\label{sec:dqpt}
The theory of DQPT relies on an intuitive connection between the thermal partition function and the Loschmidt amplitude\cite{Heyl2013}

\begin{align}\label{eq:LoschAmp}
G(t)=\bra{\psi_i}\exp(-\mathrm{i}H_ft)\ket{\psi_i},
\end{align}
where $\ket{\psi_i}$ is the initial state, $H_f$ is the quench Hamiltonian, and 
$\hbar$ is set inconsequentially to unity throughout our paper. The Loschmidt amplitude~\eqref{eq:LoschAmp} is a boundary partition function, and its logarithm is therefore proportional to a dynamical free energy:

\begin{align}\label{eq:ReturnRate}
r(t)=-\lim_{N\to\infty}\frac{1}{N}\ln|G(t)|^2,
\end{align}
where $N$ is the system size. Nonanalyticities in the thermal free energy as a function of temperature indicate critical temperatures at which thermal phase transitions occur. In~\eqref{eq:ReturnRate}, complexified time replaces the inverse temperature. Consequently, nonanalyticities of~\eqref{eq:ReturnRate} as a function of time indicate critical times at which DQPT occur.

The type of nonanalyticities that occur in the wake of a quench can either be \textit{regular} or \textit{anomalous}.\cite{Halimeh2017,Zauner2017} The former are the cusps that occur for quenches crossing the dynamical critical point, while the latter can occur for arbitrarily small quenches within the ordered phase of the system, and they have been shown to be connected to local spin excitations forming the lowest-lying excitations in the spectrum of the quench Hamiltonian.\cite{Halimeh2018a,Hashizume2018,Defenu2019}

The advantage of the LRKC is that it is integrable and also allows for a closed-form expression of its return rate, which we briefly derive here. Let us denote by $H_i$ the initial Hamiltonian in whose ground state $\ket{\psi_i}$ we prepare our system, and by $H_f$ the final Hamiltonian with which we quench. Hamiltonian $H_{i(f)}$ corresponds to~\eqref{eq:lrkc_momentum} with $h=h_{i(f)}$, where here it should be clear that we quench the chemical-potential strength. Moreover, Hamiltonians $H_{i(f)}$ are diagonalised by different Bogolioubov bases, which we denote, respectively, as $(\eta_k,\eta_k^\dagger )$ and $(\gamma_k,\gamma_k^\dagger )$, i.e.,
\begin{equation}
\begin{split}
H_{i} =& \sum_{k}^\mathrm{B.z.} \omega_k^{i}(\eta_k^\dagger \eta_k - \eta_{-k}\eta_{-k}^\dagger), \\
H_{f} =& \sum_{k}^\mathrm{B.z.} \omega_k^{f}(\gamma_k^\dagger \gamma_k - \gamma_{-k}\gamma_{-k}^\dagger) .
\end{split}
\end{equation}

The respective ground states are the vacua $\ket{0}_\eta$ and $\ket{0}_\gamma$.
For the calculation of $G(t)$ it is useful to write $\ket{0}_\eta$ in terms of $\ket{0}_\gamma$ by removing all $\eta$ fermions from $\ket{0}_\gamma$ as
\begin{equation}
\ket{0}_\eta = \mathcal{F}^{-1} \prod_k^\mathrm{ B.z.} \eta_k \ket{0}_\gamma = \mathcal{F}^{-1} \prod_{k>0} \eta_k \eta_{-k} \ket{0}_\gamma .
\end{equation}
Here $\mathcal{F}$ is a normalisation factor.

Following the BCS-theory\cite{Bardeen1957} approach one can show that
\begin{equation}\label{e:etaBCS}
\ket{0}_\eta = \mathcal{F}^{-1} \exp(-\mathrm{i}\sum_{k>0} \Lambda_k \gamma^\dagger_k \gamma^\dagger_{-k} ) \ket{0}_\gamma ,
\end{equation}
where the normalisation is given by $\mathcal{F}^2 = \prod_{k>0}(1+\Lambda_k^2)$ and $\Lambda_k = \tan(\varphi_k)$.
The angle $\varphi_k$ stems from a Bogoliubov transformation between the pre- and post-quench basis
\begin{equation}
\eta_k = \cos \varphi_k \gamma_k + \mathrm{i} \sin \varphi_k \gamma_{-k}^\dagger ,
\end{equation}
with
\begin{equation}
\varphi_k = \frac{\theta_k^{f} - \theta_k^{i}}{2} .
\end{equation}
Substituting \eqref{e:etaBCS} into \eqref{eq:LoschAmp} we obtain
\begin{equation}\label{e:G}
\begin{split}
G(t) =&\, \mathcal{F}^{-2} \prescript{}{\gamma}{\bra{0}}  \exp\Big(+\mathrm{i}\sum_{k>0} \Lambda_k \gamma_{-k} \gamma_{k} \Big) e^{-\mathrm{i} H_f t} \\
&\times \exp\Big(-\mathrm{i}\sum_{k>0} \Lambda_k \gamma^\dagger_k \gamma^\dagger_{-k} \Big)\ket{0}_\gamma .
\end{split}
\end{equation}
Using the unitarity of $e^{-\mathrm{i} H_f t}$, we write the product of the last two exponential terms above as
\begin{equation}\label{e:prod}
\begin{split}
&\exp\Big(-\mathrm{i}\sum_{k>0}\Lambda_k e^{-\mathrm{i}tH_f} \gamma^\dagger_k \gamma^\dagger_{-k} e^{+\mathrm{i}tH_f}\Big) e^{-\mathrm{i}tH_f} \\
=& \exp\Big(-\mathrm{i}\sum_{k>0}\Lambda_k \gamma^\dagger_k \gamma^\dagger_{-k} e^{-\mathrm{i}t 2 \epsilon_k^{f}}\Big) e^{-\mathrm{i}tH_f},
\end{split}
\end{equation}
where we have defined $\epsilon_k^{f} = 2 \omega_k^{f}$. To arrive at the second line we have used that
\begin{equation}
e^{-\mathrm{i}tH_f} \gamma^\dagger_{\pm k} e^{\mathrm{i}tH_f} =  \gamma_{\pm k}^\dagger e^{-\mathrm{i}t 2\omega_k^{f}} ,
\end{equation}
which follows from $\omega_k^{f} = \omega_{-k}^{f}$.

Substituting \eqref{e:prod} back into \eqref{e:G}, and applying Pauli exclusion to the Taylor expansion of the remaining exponential terms, then yields
\begin{equation}
\begin{split}
G(t) =& \mathcal{F}^{-2} \prescript{}{\gamma}{\bra{0}} \prod_{k>0} (1 + \mathrm{i} \Lambda_k \gamma_{-k} \gamma_{k}) \\
&\times (1 - \mathrm{i} \Lambda_k \gamma_{k}^\dagger \gamma_{-k}^\dagger e^{-\mathrm{i}t 2 \epsilon_k^{f}}) \ket{0}_\gamma ,
\end{split}
\end{equation}
where we have dropped an inconsequential phase arising from the $e^{-\mathrm{i}tH_f} \ket{0}_\gamma$ term. Taking care of the commutation relations when multiplying out the above product, and writing the remaining terms in normal ordered form, allows us to simplify the expression to
\begin{equation}
G(t) = \mathcal{F}^{-2} \prod_{k>0} (1 + \Lambda_k^2 e^{-\mathrm{i}t 2 \epsilon_k^{f}}) .
\end{equation}
It is then straightforward to show that the return rate is
\begin{equation}
\begin{split}\label{eq:RR}
r(t)=&-\lim_{N\to\infty}\frac{1}{N} \sum_{k>0} \ln\big[1-\sin^2(\theta_k^{f} - \theta_k^{i}) \sin^2(\epsilon_k^{f}t)\big] \\
=&-\int_0^\pi \frac{dk}{2\pi} \ln\big[1-\sin^2(\theta_k^{f} - \theta_k^{i}) \sin^2(2 \omega_k^{f}t)\big] .
\end{split}
\end{equation}\\

Appendix~\ref{app:compare} provides a comparison between the return rate~\eqref{eq:RR} in the limit of $\alpha,\beta\to\infty$ and its counterpart in the TFIC

\begin{align}\label{eq:Ising}
H_\text{Ising}=-\sum_j\big[\sigma^z_j\sigma^z_{j+1}+h\sigma^x_j\big].
\end{align} The latter exactly maps onto the LRKC~\eqref{eq:lrkc_momentum} in this limit. Immediately one sees that~\eqref{eq:RR} can show nonanalytic behavior only in the momentum sectors $k_c$ that satisfy $\theta_{k_c}^f-\theta_{k_c}^i=\pi/2+n\pi$, with $n\in\mathbb{Z}$, or, in other words, the mode $k_c$ is in an infinite-temperature state. One can further show that these critical momenta satisfy the implicit relation

\begin{align}\label{eq:critmom}
h_ih_f-J_{k_c}(h_i+h_f)+J_{k_c}^2+\Delta_{k_c}^2=0,
\end{align}
which in the case of nearest-neighbor couplings \mbox{($\alpha,\beta\to\infty$)}, reduces to

\begin{align}\label{eq:critmomNN}
k_c^\text{NN}=\arccos\frac{\big(h_c^e\big)^2+h_i h_f}{h_c^e(h_i+h_f)}.
\end{align}
As we shall discuss in great detail later, the relation~\eqref{eq:critmom} is satisfied if and only if $h_i$ and $h_f$ are on opposite sides of the dynamical critical point $h_c^d\leq h_c^e$. Only quenches where the value of $h_f$ leads to at least one real solution of~\eqref{eq:critmom} can therefore lead to nonanalytic behavior in~\eqref{eq:RR}. When~\eqref{eq:critmom} is satisfied, one finds nonanalytic behavior in~\eqref{eq:RR} at the critical times

\begin{align}
t_n^*=\bigg(n+\frac{1}{2}\bigg)\frac{\pi}{2 \omega_{k_c}^f},\,\,\,\,\,\,n\in\mathbb{N}.
\end{align}
The TFIC can be mapped using a Jordan-Wigner transformation onto the LRKC model~\eqref{eq:lrkc_momentum} in the limit $\alpha,\beta\to\infty$. In this case,~\eqref{eq:critmom} reduces to~\eqref{eq:critmomNN} and yields a real unique value of $k_c$ if and only if $h_i$ and $h_f$ are on different sides of the quantum equilibrium critical point $h_c^e$, which in this case equals $h_c^d$. In the case of the general LRKC model, the situation is more subtle in that cusps from the topological phase to just above $h_c^e$ can give rise to three critical momenta;\cite{Dutta2017} cf.~Sec.~\ref{sec:DPD}. More intriguingly, for certain regimes of $(\alpha,\beta)$, which we discuss in detail in Sec.~\ref{sec:DPD}, doubly critical cusps~---~which are related to two distinct critical momenta~---~can arise for quenches within the topologically nontrivial phase.\cite{Defenu2019} As we elaborate in Sec.~\ref{sec:results}, these cusps may explain the so-called \textit{accidental} dynamical vortices observed in the experiment of Ref.~\onlinecite{Flaeschner2018}. It is worth noting here that LRKC has also been used to study the effect of edge modes on the boundary contributions to the DQPTs and the dynamics of entanglement entropy.\cite{Sedlmayr18}

\section{Dynamical phase diagram}\label{sec:DPD}
Now we present one of the main results of our work, which is an analytic derivation of the dynamical phase diagram further augmented by numerical results.

In equilibrium, long-range interactions do not cause the appearance of any additional phases, but rather alter the value of the critical boundaries. Critical points are characterised by the occurrence of massless excitations and, then, they can be identified by locating the zeros of the single particle spectrum in \eqref{e:Ham2} . The condition $\Delta_{k}=0$ identifies two critical modes $k=0,\pi$, each yielding a critical point respectively at $h=h^{e}_{c}=(1,-1+2^{1-\alpha})$; as can be deduced from the value of $J_{k=0,\pi}$, see \eqref{eq:Jk} . These boundaries reduce to the nearest-neighbor case $h^{e}_{c}=(1,-1)$ in the $\alpha\to\infty$  limit, irrespective of the value of $\beta$ since the condition $\Delta_{k}=0$ always holds for $k=0,\pi$.

Conversely, the presence of two different power-law decays for the hopping and pairing matrices produces a rich dynamical phase diagram in the LRKC~\eqref{eq:lrkc}, with novel dynamical phases that have no equilibrium counterparts. As already established,~\cite{Heyl_review} quenching the chemical potential $h$ across one of the equilibrium critical points produces a dynamical phase transition, which takes the form of regular\cite{Halimeh2017} singularities in the system's return rate. In the TFIC the origin of these singularities can be traced back to the appearance of a perfectly athermal mode whose excitation probability is exactly $p_k=1/2$. For simplicity and without loss of qualitative generality, we shall henceforth restrict ourselves to nonnegative values of the chemical potential.

Conversely, different situations arise for general $\alpha$ and $\beta$ where more than a single dynamical critical mode can occur for specific quenches. It has already been noticed that in the case of nearest-neighbor hopping and long-range pairing a triply critical phase can emerge with three excitation modes having $p_{k}=1/2$ for quenches across the topological phase transition.\cite{Dutta2017} More recently, the occurrence of a doubly critical dynamical phase in which two critical modes arise has been demonstrated for the case of long-range pairing and hopping with equal decay rate $\beta=\alpha$ for quenches within the topologically nontrivial phase, i.e., from $h_i<h_c^e$ to $h_f\in(h_c^d,h_c^e)$.\cite{Defenu2019}

It is worth noting that the properties of these peculiar dynamical phases can hardly be connected with those of equilibrium critical phenomena. Indeed, long-range couplings alter the universal behavior close to the equilibrium critical points only for $\alpha<2$ or $\beta<2$. One can immediately see this by considering the equilibrium dynamical critical exponent $z$,
\begin{align}
\label{dyn_crit_exp}
z=\begin{cases}
\phi-1&\quad\mathrm{if}\,\, \phi < 2,\\
1 &\quad\mathrm{if}\,\, \phi \geq 2,
\end{cases}
\end{align}
where $\phi=\mathrm{min}(\alpha,\beta)$ and we have used the definition ${\omega_{k}\propto\,k^{z}}$ at $h=h^{e}_{c}$. The product between the dynamical critical exponent $z$ and the correlation length exponent $\nu$ always remains the same as in the nearest-neighbor interacting case, i.e., $z\nu=1$. Note that the relevant region for long-range couplings in the LRKC is radically different with respect to the case of interacting field theories.\cite{Defenu2017} On the other hand, the doubly critical phase also arises in the case of irrelevant long-range couplings $\phi\geq 2$, demonstrating that the dynamical critical behavior of long-range systems is not as strictly tied to the equilibrium universality as in nearest-neighbor systems.\cite{Homrighausen2017} A similar discrepancy in the relevance of long-range couplings, first noticed in Ref.~\onlinecite{Defenu2019} for the $\beta=\alpha$ case, also exists in the present general case. 

\begin{figure*}[!ht]
	\centering
	\vspace{-.1 cm}
	\includegraphics[width=.45\textwidth]{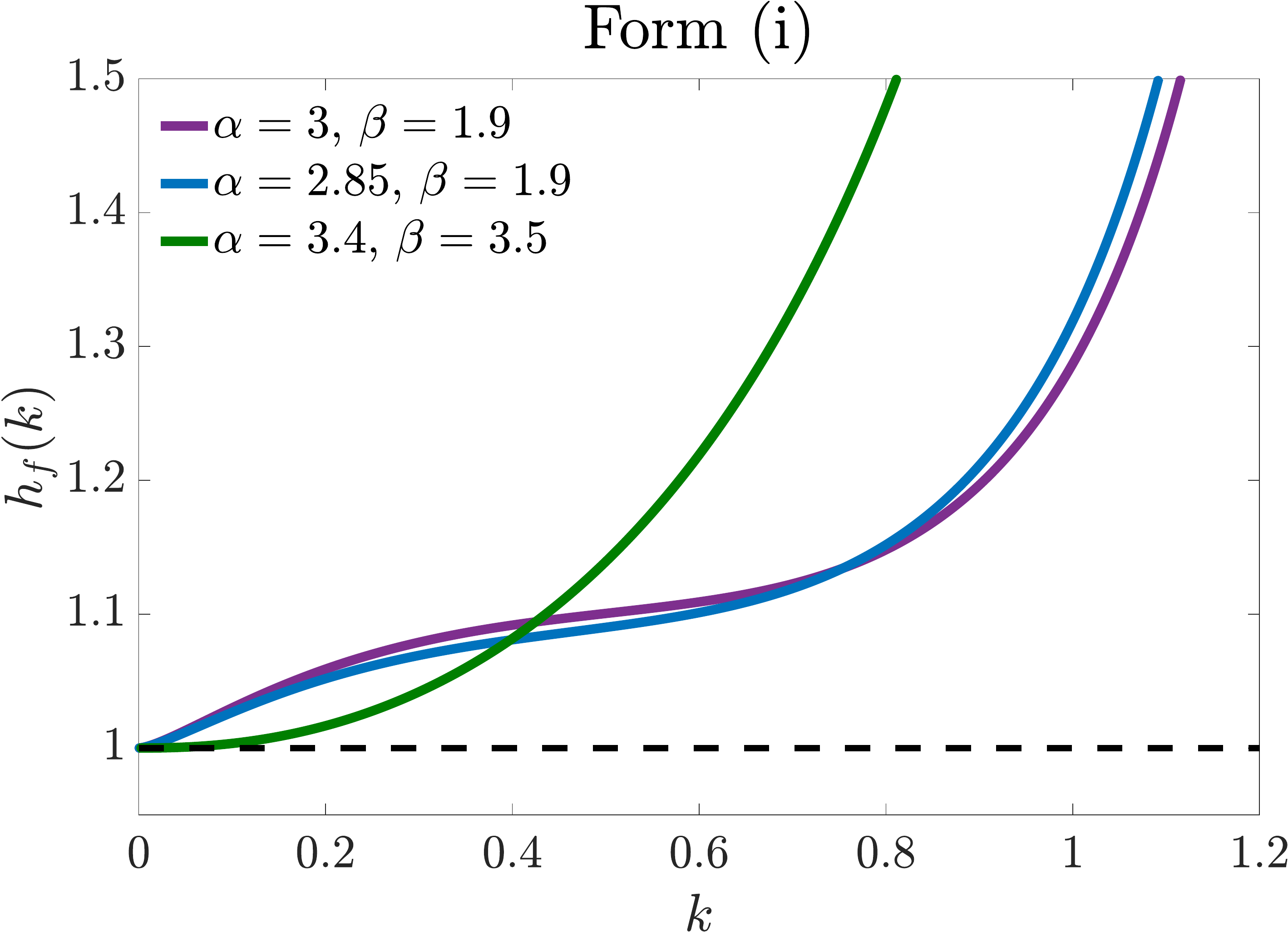}\quad
	\vspace{-.01 cm}
	\includegraphics[width=.45\textwidth]{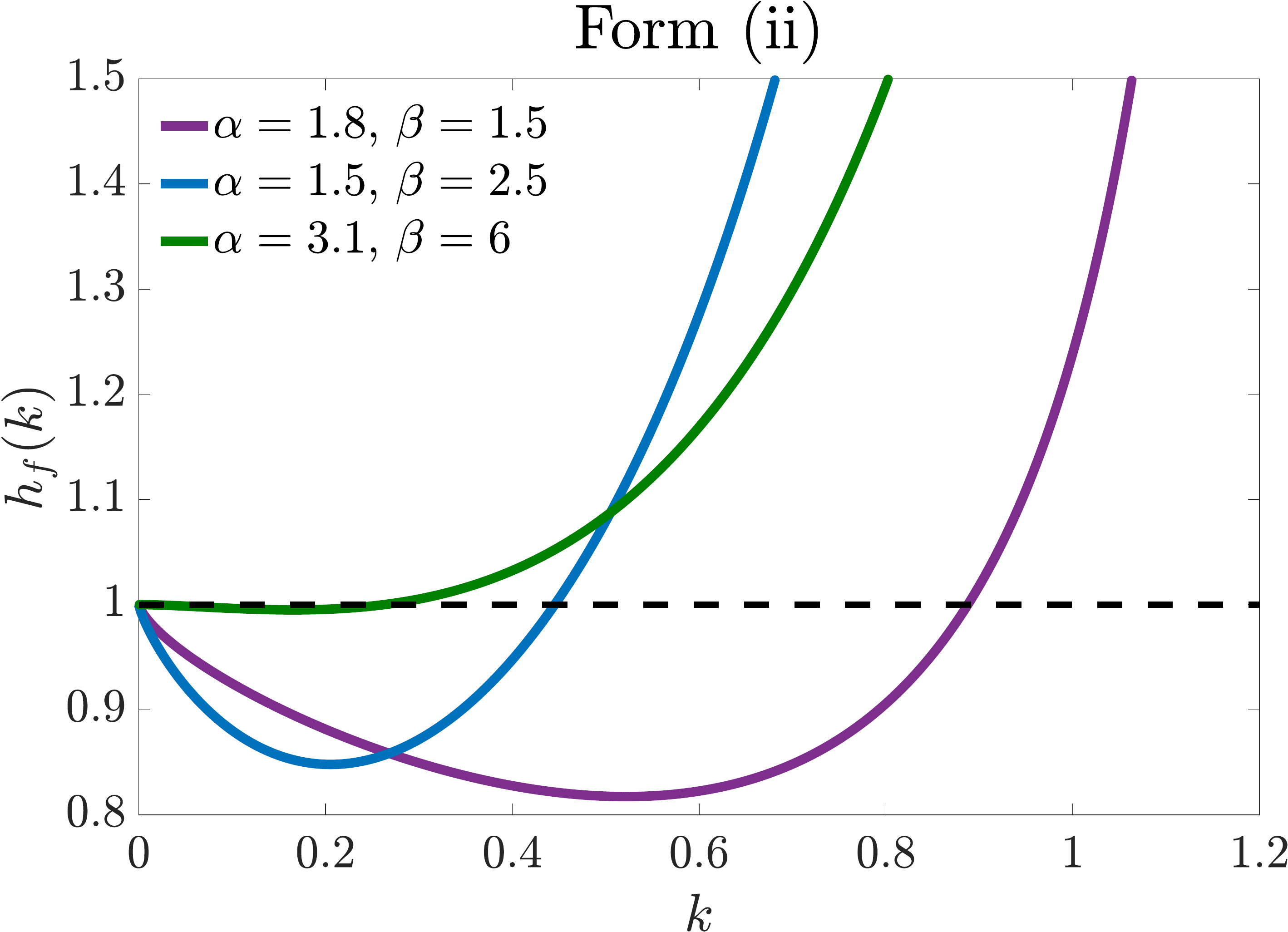}\quad
	\vspace{.05 cm}\\
	\hspace{-.5 cm}
	\includegraphics[width=.45\textwidth]{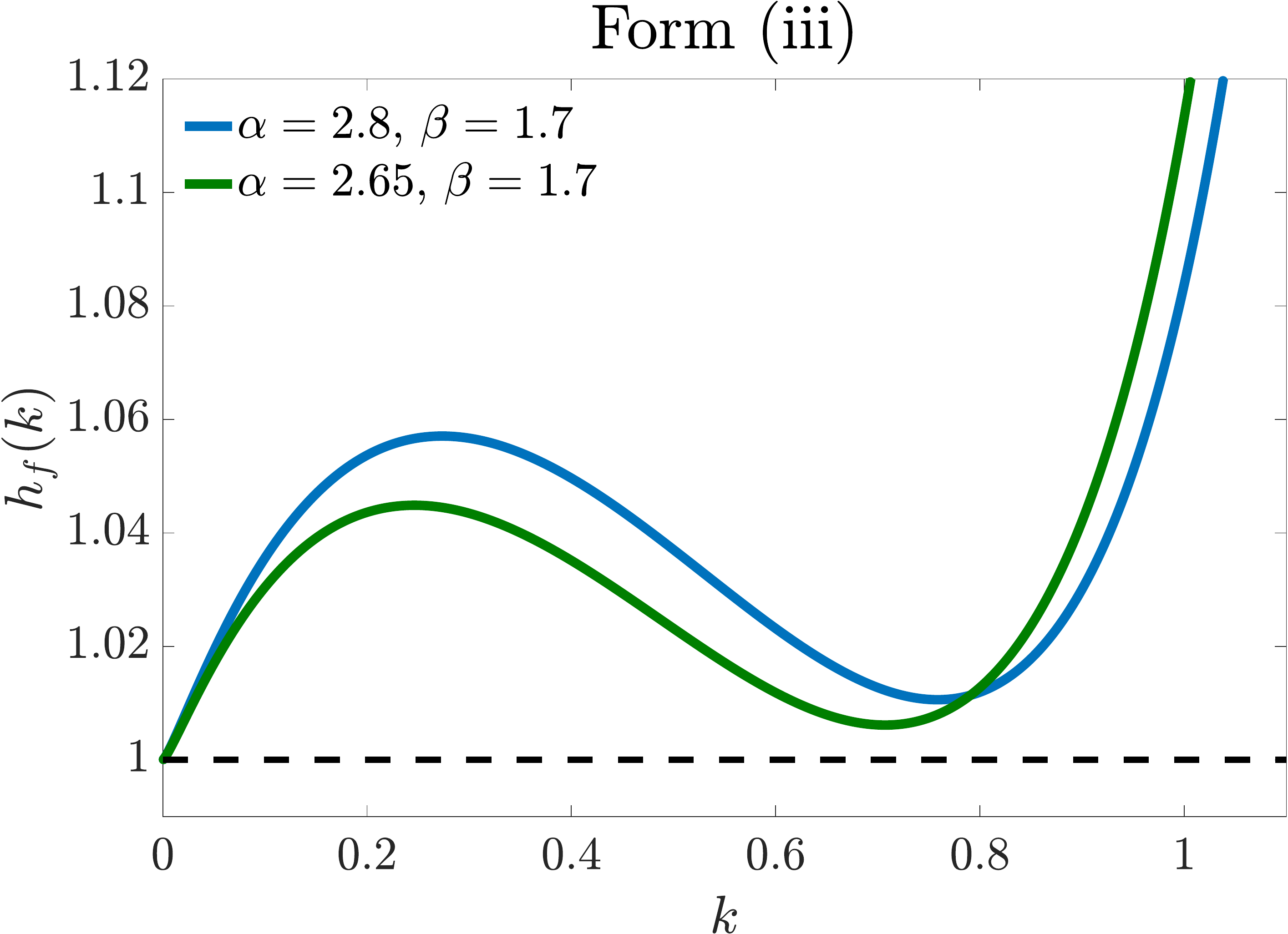}\quad
	\hspace{.01 cm}
	\includegraphics[width=.444\textwidth]{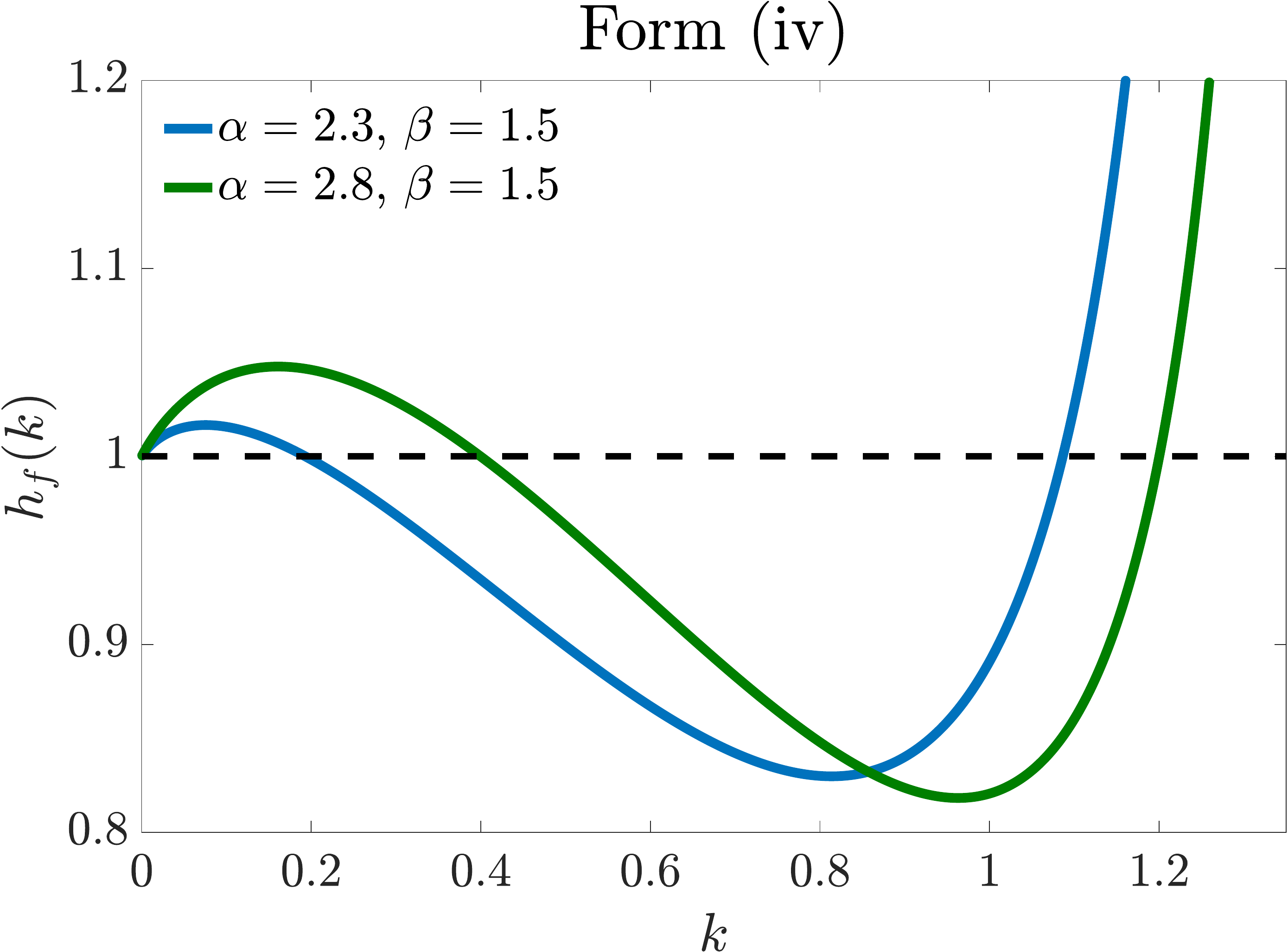}
	\hspace{-.01 cm}
	\caption{(Color online). The four different forms that function~\eqref{eq:hf_reduced} can take for quenches from $h_i=0$. In form (i), we see that there exist only the singly critical phase, where quenches to $h_f>h_c^e=1$ generate a unique critical momentum mode that gives rise to critical times in the return rate. In form (ii), we see again the singly critical phase, but also the doubly critical phase, which occurs for quenches within the topologically nontrivial phase and allows for two critical momentum modes. In form (iii), the singly critical phase emerges for quenches just above $h_c^e$ and also for large quenches. Quenches in between lead to a triply critical phase with three critical momentum modes. Finally, form (iv) exhibits the richest dynamical criticality, in that again the doubly critical phase emerges for quenches within the topologically nontrivial phase, the triply critical phase appears for quenches in a region above $h_c^e$, but for quenches above that region only the singly critical phase exists.} 
	\label{fig:DPD} 
\end{figure*}

After a quench of the chemical potential $h$ in the model~\eqref{eq:lrkc}, quasiparticles in certain momentum sectors $k$ may attain infinite temperature with excitation probability $p_{k}=1/2$.
The value of the final chemical potential $h_{f}$ at which the momentum $k$ becomes critical after a quench starting at the chemical potential $h_{i}$ is given by\cite{Defenu2019}
\begin{align}\label{eq:hf}
h_{f}(k)=J_k+\frac{\Delta_k^{2}}{J_k-h_{i}},
\end{align}
which is a mere rearrangement of~\eqref{eq:critmom},
with $J_k$ and $\Delta_k$ given in~\eqref{eq:Jk} and~\eqref{eq:Deltak}.
For simplicity, and without loss of qualitative generality, we shall restrict our analysis to quenches starting at $h_{i}=0$, which simplifies~\eqref{eq:hf} to
\begin{align}\label{eq:hf_reduced}
h_{f}(k)=J_k+\frac{\Delta_k^{2}}{J_k}.
\end{align}
Physically, the relation between the dynamical and equilibrium phase diagrams is due to the peculiarity of the $k=0$ mode. Indeed, the pairing gap in the Hamiltonian in \eqref{eq:lrkc_momentum} vanishes at $k=0$, and the $k\approx 0$ modes decouple from the dynamics since they are eigenstates of the momentum density operator and cannot be excited by a time-dependent chemical potential, at least so long as the equilibrium critical boundary is not crossed. Yet, if the chemical potential $h$ dynamically crosses the equilibrium threshold $h^{e}_{c}$, the $k=0$ ground state changes its nature, but not its occupation, leading to maximal excitation probability $p_{k=0}=1$ for $h_{f}>h^{e}_{c}$. Moreover, quenches between $h_{i}=0$ and $h_{f}>0$ never cause excitations beyond the critical threshold ($p_{k}=1/2$) for high-energy modes ($k>\pi/2$).

Therefore, when $h_{f}(k)$ has a single minimum, which lies below the equilibrium critical threshold ($\min_{k}h_{f}(k)=h_{\mathrm{min}}<h^{e}_{c}$), all the quenches with  $h_{\mathrm{min}}<h_{f}<h^{e}_{c}$ develop two critical modes, as the excitation probability shall cross the critical value $p_{k}=1/2$ twice in order to satisfy the boundary $p_{k=0}=0$ and $p_{k}<1/2$ at large $k$. Using the same argument one may derive the whole dynamical phase diagram by a careful study of the function $h_{f}(k)$. The four possible different \textit{forms} of $h_f(k)$ are shown in Fig.~\ref{fig:DPD}:
\begin{enumerate}[(i).]
	\item The function $h_{f}(k)$ is monotonous and the system has only the traditional \textit{singly} critical dynamical phase with $h_{c}^{d}=h_c^e=1$ with just a single critical momentum.\cite{Heyl2013}
	\item The function $h_{f}(k)$ has a single minimum with value $h_{c}^{d}<1$ such that for $h_{c}^{d}<h_{f}<1$ a doubly critical phase with two critical momenta emerges, whereas when $h_f>1$ a singly critical phase with a single critical mode appears. In this case, $h_f(k)$ has no local maximum.
	\item The function $h_{f}(k)$ has a local minimum and a local maximum both at $k>0$, where both the maximum and the minimum lie in the region $h_{f}>1$ such that $h_{\text{max}}>h_{\text{min}}>1$. This gives rise to a singly critical phase for $1<h_{f}<h_{\text{min}}$ or $h_{f}>h_{\text{max}}$, and a triply critical phase, which has three critical momentum modes, for $h_{\text{min}}<h_{f}<h_{\text{max}}$. In this case, the local maximum occurs at a smaller momentum relative to the local minimum.
	\item The local minimum has a value $h_\text{min}=h_{c}^{d}<1$, while the local maximum always occurs at $h_{\text{max}}>1$, leading to the emergence of all three dynamically critical phases: The doubly critical phase for $h_c^d<h_f<1$, the triply critical phase for $1<h_f<h_\text{max}$, and the singly critical phase for $h_f>h_\text{max}$. Also in this case, the local maximum appears at a smaller momentum relative to the local minimum.
\end{enumerate}

These forms show that the triply critical phase can only emerge at $h_{f}>1$, and that the doubly critical phase only appears for $h_{f}<1$. In the case of form (ii) where only a local minimum, but not a local maximum, exists, the doubly critical phase can be detected by the sign of the leading correction to the function $h_{f}(k)$ at $k\simeq0$, while for form (iv) the Taylor expansion around $k=0$ cannot be used to detect the doubly critical phase. In short, when the sign of the leading correction to $h_f(k)$ is negative, one can be certain that form (ii) arises, whereas for a positive sign any of the other three forms can occur.

In the following, our approach to uncover most of the properties of the dynamical phase diagram will be to study the function~\eqref{eq:hf_reduced} around $k=0$. To ease the analysis, we split $h_{f}(k)$ into two contributions
\begin{align}\nonumber
h_1=J_k=&\,1+\frac{\Gamma(1-\alpha)\sin(\pi\alpha/2)}{\zeta(\alpha)}k^{\alpha-1}\\
&-\frac{\zeta(\alpha-2)}{2\zeta(\alpha)}k^{2}+\order{k^{3}},\\\nonumber
h_2=\frac{\Delta_k^{2}}{J_k}=&\left[\frac{\Gamma(1-\beta)}{\zeta(\beta)}\cos\frac{\beta\pi}{2}\,k^{\beta-1}+\frac{\zeta(\beta-1)}{\zeta(\beta)}k\right]^{2}\\
&+\order{k^{3}}.
\end{align} 
Conveniently, the two exponents $\alpha$ and $\beta$ enter separately into these contributions. This allows for an independent analysis of their influence on the forms (i)--(iv).
\subsection{The leading long-range pairing case ($\beta<2$)}
In the case $\beta<2$ the second contribution is dominated by the nonanalytic power, leading to
\begin{align}
h_2=\frac{\Gamma^2(1-\beta)}{\zeta^2(\beta)}\cos^2\frac{\beta\pi}{2}\,k^{2\beta-2}+\order{k^\beta}.
\end{align}
Consequently, only two possibilities can arise, as follows.
\subsubsection{The $\alpha<(2\beta-1)$ regime}
In this regime the leading-order expansion for $h_{f}(k)$ is given by $h_1$ as
\begin{align}
h_{f}(k)=1+\frac{\Gamma(1-\alpha)\sin(\pi\alpha/2)}{\zeta(\alpha)}k^{\alpha-1}+\order{k^{2\beta-2}}.
\end{align}
The coefficient $\Gamma(1-\alpha)\sin(\pi\alpha/2)/\zeta(\alpha)$
is always negative and therefore the doubly critical phase is always present in this regime. The subleading term is given by $h_2$ and is always positive. In this regime there is no triply critical phase. Examples of this regime are shown in the top right panel of Fig.~\ref{fig:DPD}.

\subsubsection{The $\alpha>(2\beta-1)$ regime}
In this case the roles of $\alpha$ and $\beta$ are reversed. The leading-order term is given by $h_2$ and the expansion reads
\begin{align}
h_{f}(k)=1+\frac{\Gamma^2(1-\beta)}{\zeta^2(\beta)}\cos^2\frac{\beta\pi}{2}\,k^{2\beta-2}+\cdots.
\end{align}
In this regime, the leading term is always positive and thus the existence of the doubly critical phase cannot be directly inferred. The $\alpha,\beta$ dependence of the subleading correction depends on the value of $\alpha$.

For ${\alpha<\beta+1}$, the subleading term is ${[\Gamma(1-\alpha)\sin(\pi\alpha/2)/\zeta(\alpha)]k^{\alpha-1}}$, whereas for ${\alpha>\beta+1}$ it is ${2[\Gamma(1-\beta)\zeta(\beta-1)\cos(\pi\beta/2)/\zeta^2(\beta)]k^{\beta}}$. In either case the slope is negative, implying that the subleading term can overcome the initial positive slope of the leading term at $h_f(k=0)=1$. Consequently, $h_f(k)$ can go below unity, in which case one can be sure of the existence of the doubly critical phase, i.e., form (iv).

In such a case, the triply critical phase exists. Examples for this are shown in the bottom right panel of Fig.~\ref{fig:DPD}, for $\alpha<\beta+1$ ($\alpha=2.3$ and $\beta=1.5$) and for $\alpha>\beta+1$ ($\alpha=2.8$ and $\beta=1.5$), where in both cases $\alpha>2\beta-1$. 
	
However, it is also possible that form (i) or (iii) may emerge instead, in which no doubly critical phase exists as both the maximum and minimum of $h_f(k)$ occur above unity. As examples, we show in the top left panel of Fig.~\ref{fig:DPD} how for $\alpha<\beta+1$ ($\alpha=2.85$ and $\beta=1.9$) and for $\alpha>\beta+1$ ($\alpha=3$ and $\beta=1.9$) form (i) can emerge where only a singly critical phase is possible. Additionally, we shown examples of how form (iii) may arise in the bottom left panel of Fig.~\ref{fig:DPD} for $\alpha<\beta+1$ ($\alpha=2.65$ and $\beta=1.7$) and for $\alpha>\beta+1$ ($\alpha=2.8$ and $\beta=1.7$). Note that all these examples satisfy $\alpha>2\beta-1$.

\subsection{The subleading long-range pairing case ($\beta>2$)}\label{sec:triply}
When $\beta>2$, the second contribution to $h_f(k)$ is dominated by the analytic power,
\begin{align}
h_2=\frac{\zeta^2(\beta-1)}{\zeta^2(\beta)}k^{2}+\order{k^{2\beta-2}} ,
\end{align}
which leads to two different regimes as a function of $\alpha$. It is worth noting that in this region no triply critical phase can ever exist as we shall discuss in the following.
\subsubsection{The $\alpha<3$ regime}
In this case the Taylor expansion for the critical final field reads
\begin{align}\label{e:alpha<3}
h_{f}(k)=1+\frac{\Gamma(1-\alpha)\sin(\pi\alpha/2)}{\zeta(\alpha)}k^{\alpha-1}+\order{k^{2}}.
\end{align}
Since the coefficient is always negative, the doubly critical phase is always present in this regime. This would bring about form (ii), which excludes the emergence of a triply critical dynamical phase. An example is shown in the top right panel of Fig.~\ref{fig:DPD} for $\alpha=1.5$ and $\beta=2.5$.

\subsubsection{The $\alpha>3$ regime}
\begin{figure}[!ht]
	\centering
	\hspace{-.25 cm}
	\includegraphics[width=.4\textwidth]{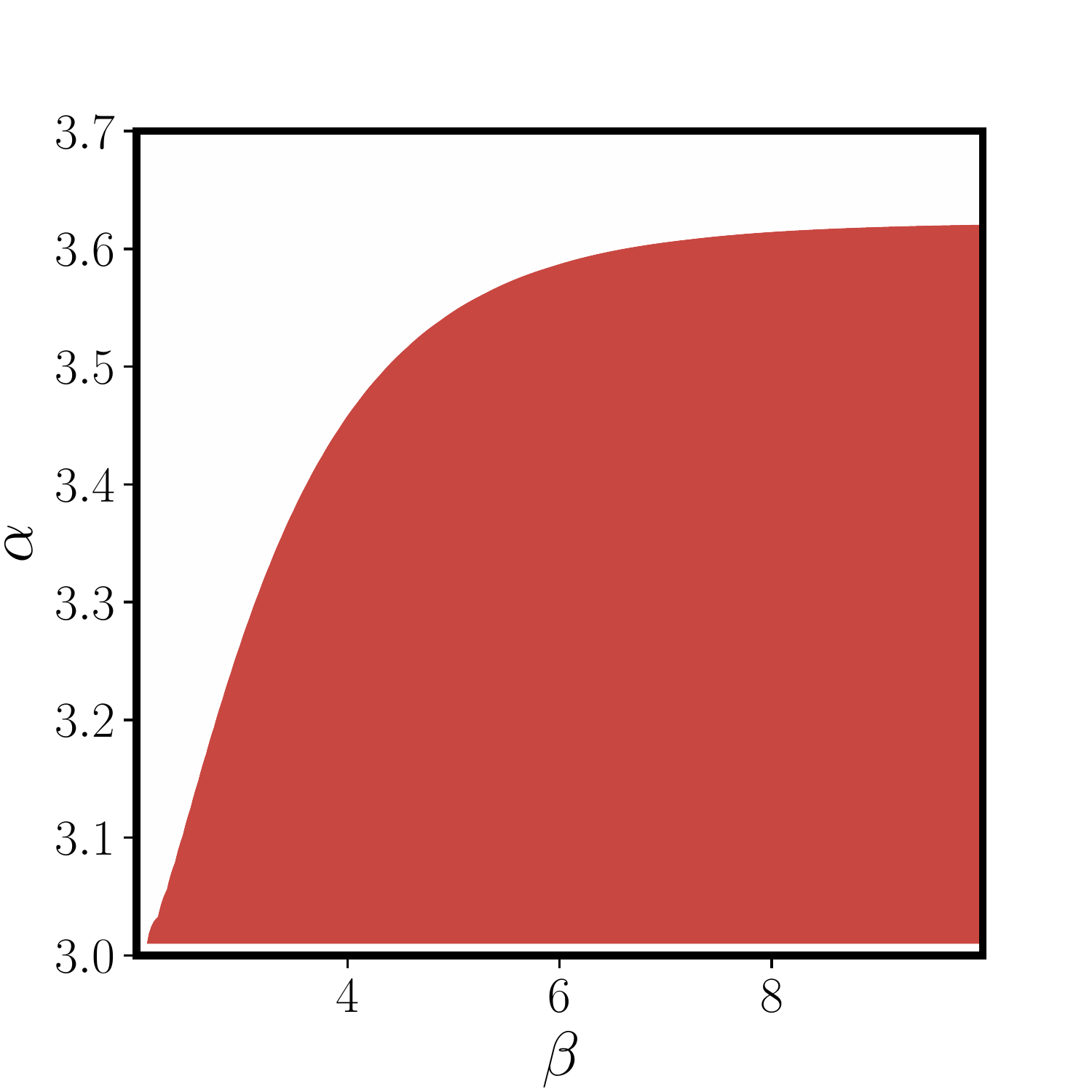}\quad
	\hspace{-.15 cm}
	\caption{(Color online). Phase diagram of the model in the $(\alpha,\beta)$ plane. In the red-colored region the doubly critical phase exists, while in the white region $c_1>0$ and there is no doubly critical phase.} 
	\label{Fig5} 
\end{figure}

In this regime both of the nonanalytic terms are irrelevant and the leading contribution reads
\begin{align}
h_{f}(k)=1+\left[\frac{\zeta^2(\beta-1)}{\zeta^2(\beta)}-\frac{\zeta(\alpha-2)}{2\zeta(\alpha)}\right]k^{2}+\order{k^{2\beta-2}}.
\label{c1}
\end{align}
The existence of the doubly critical phase depends on the coefficient 
\begin{align}
c_{1}=\frac{\zeta^2(\beta-1)}{\zeta^2(\beta)}-\frac{\zeta(\alpha-2)}{2\zeta(\alpha)}.
\end{align}
For $c_{1}<0$ the doubly critical phase exists within form (ii), meaning that the triply critical phase cannot exist in this case, and an example of this regime is given in the top right panel of Fig.~\ref{fig:DPD} for $\alpha=3.1$ and $\beta=6$. On the other hand, for $c_{1}>0$ the function $h_{f}(k)$ is monotonous and only the singly critical phase exists, as illustrated by the example for $\alpha=3.4$ and $\beta=3.5$ shown in the top left panel of Fig.~\ref{fig:DPD}. The dynamical phase diagram of the model in this region is shown in Fig.~\ref{Fig5}.

\begin{figure}[!ht]
	\centering
	\hspace{-.25 cm}
	\includegraphics[width=.43\textwidth]{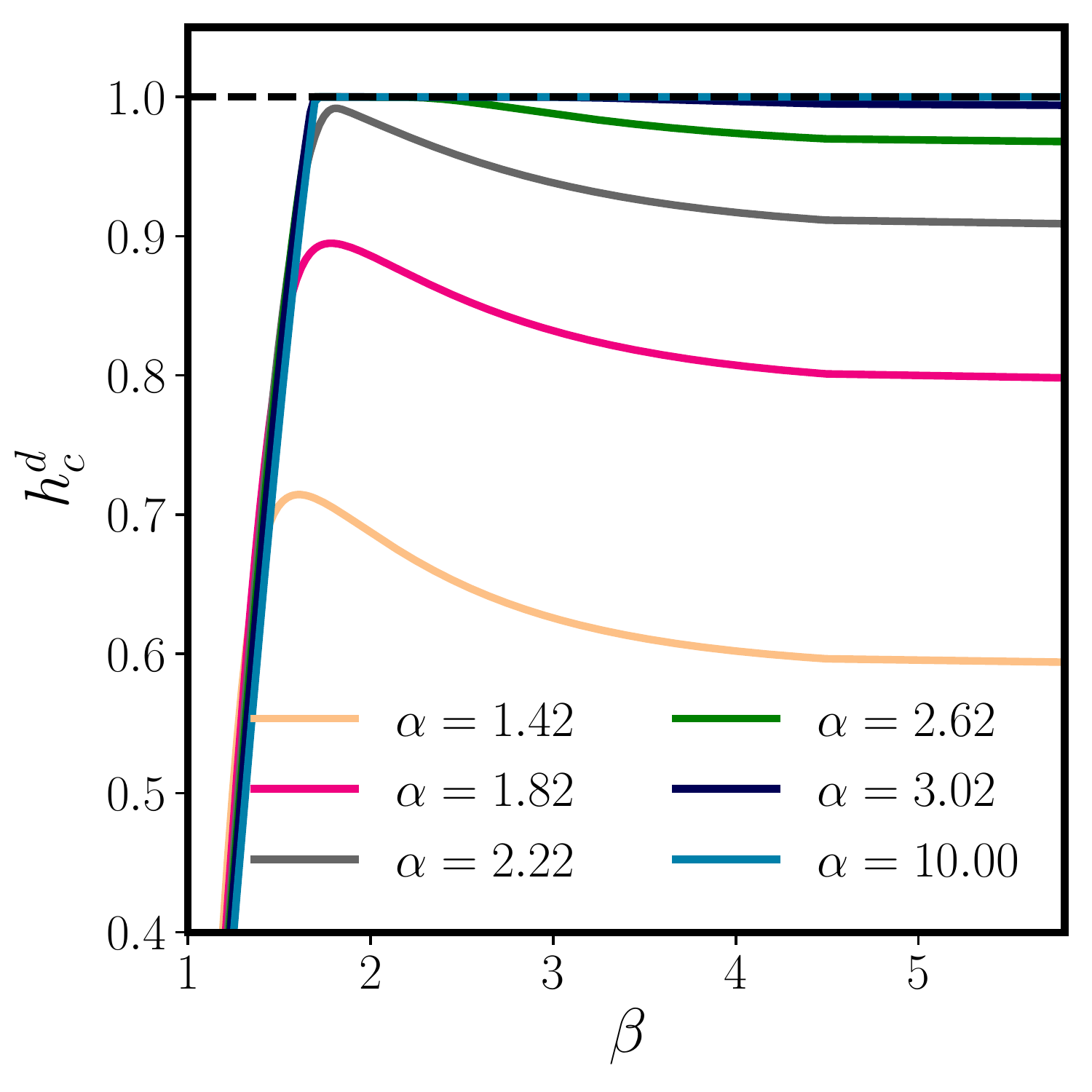}\quad
	\hspace{-.15 cm}
	\caption{(Color online). The value $h_{c}^{d}$ which delimits the appearance of the doubly critical phase for a quench starting at $h_{i}=0$ is shown as a function of $\beta$ for different values of $\alpha$.} 
	\label{Fig2} 
\end{figure}

As we discuss in Sec.~\ref{sec:results}, of particular interest to us is the doubly critical phase. This is due to recent experimental observations\cite{Flaeschner2018} of so-called \textit{accidental} dynamical vortices that appear in time-resolved Bloch tomography measurements for quenches within the same topological phase. Our results indicate that this is indeed possible even in models with no interactions such as the Haldane-like model realized in Ref.~\onlinecite{Flaeschner2018}. Indeed, the generally accepted explanation~---~ based on results of long-range interacting quantum spin models~---~for why the dynamical critical point can be smaller than the equilibrium critical point has been that this is most likely due to interactions.\cite{Halimeh2017,Halimeh2017b,Zauner2017,Homrighausen2017,Lang2017,Lang2018,Halimeh2018a,Hashizume2018,Mori2019} Nevertheless, as indicated in our previous work Ref.~\onlinecite{Defenu2019} and solidified in our current study, the case of $h_c^d<h_c^e$ is very prevalent for noninteracting free fermionic systems with long-range hopping and pairing. Fig.~\ref{Fig2} shows $h_c^d$ as a function of the pairing exponent $\beta$ at fixed values of the hopping exponent $\alpha$. For small to intermediate values of $\alpha$ this function is not monotonous, but instead goes from very small values at $\beta\gtrsim1$, to a maximum (which is still below unity) at intermediate values of $\beta$, and then asymptotically towards a smaller value at $\beta\to\infty$. 

The case of nearest-neighbor hopping ($\alpha\to\infty$) and pairing ($\beta\to\infty$) deserves special attention. A sketch of the dynamical phase diagram for the triply critical phase in the $\alpha\to\infty$ limit has been obtained numerically in Ref.~\onlinecite{Dutta2017}, but no doubly critical phase has been found, consistently with the results depicted in Fig.~\ref{Fig5}, where the doubly critical phase disappears at large $\alpha$. On the other hand, the purely long-range hopping case $\beta\to\infty$ always contains a doubly critical phase for $\alpha\lesssim3.6223$, as follows from \eqref{e:alpha<3} and \eqref{c1}, but it does not display the triply critical phase for any $\alpha$ value.

\section{String order parameter}\label{sec:sop}

The nonlocal form of the SOP makes it a useful quantity to use when studying transitions between equilibrium topological phases. den Nijs and Rommelse first showed that the SOP witnesses the hidden N\'eel order of the Haldane phase of the spin-$1$ AKLT model.\cite{denNijsRommelse} More recent work has extended the use of SOPs in the analysis of topological phases also to bosonic and fermionic systems.\cite{Bahovadinov19, Anfuso07, Catuneanu19, Xu18, Lisandrini17} Recently, within the context of high $T_c$ superconductivity, SOPs have been used to theoretically study the topological phases diagram of the $t-J$ model.\cite{Fazzini19} It has further been proposed to study this model in quantum gas microscopes,\cite{Montorsi19} as they are ideal platforms for the measurement of SOPs.\cite{Endres2011,Hilker17}

Since the return rate is by definition also nonlocal, it is natural to ask whether the dynamical evolution of the SOP after a quench captures physics similar to that captured by the return rate. The recent work of Budich and Heyl \cite{Budich2016} has indeed shown a link between these two quantities for a Kitaev chain with next-to-nearest-neighbor hopping and nearest-neighbor pairing interactions; For a quench from the topologically nontrivial phase into the trivial phase, the authors showed that the times at which zero crossings of the SOP occur are exactly the critical times of the return rate. As a natural extension of this work, we examine in the subsequent section whether this correspondence also holds in the general Kitaev chain with long-range hopping and pairing \eqref{eq:lrkc}, thereby extending the link between DQPT and the underlying equilibrium phases to this model. This is especially interesting because we know from quantum Ising chains that the presence of long-range interactions leads to a more subtle picture in the connection between DQPT and the longitudinal magnetization. For instance, for sufficiently small quenches, anomalous cusps appear even when the order parameter never changes sign. Also the regular cusps~---~which are connected to zero crossings of the order parameter~---~display slightly different periodicities compared to the cusps of the return rate at short to intermediate evolution times, as shown, for example, in the case of the Lipkin-Meshkov-Glick (LMG) model.\cite{Homrighausen2017,Lang2017}

In this section we show how the SOP can be calculated numerically for this model.\cite{note} We write the SOP defined in Ref.~\onlinecite{Budich2016} as
\begin{equation}\label{e:sop1}
O_{lm}(t) = \phi_l^-(t) \left( \prod_{j=l+1}^{m-1} \phi_j^+(t) \phi_j^-(t) \right) \phi_m^+(t) ,
\end{equation}
where $\phi_i^\pm = c_i^\dagger \pm c_i$ and the time evolution is generated by the post-quench Hamiltonian $H_f$. Our aim is to numerically calculate
\begin{equation}\label{e:sop2}
\mathcal{M}(t) \coloneqq \lvert\! \mel{\psi_i}{O_{lm}(t)}{\psi_i} \rvert,
\end{equation}
i.e., the absolute value of the expectation value of \eqref{e:sop1} with respect to the ground state of the pre-quench Hamiltonian $H_i$. Here we give a brief summary of how this is achieved.

First, since \eqref{e:sop2} is defined with respect to $\ket{\psi_i}$, it is useful to write \eqref{e:sop1} in normal-ordered form via Wick's theorem. A direct consequence of this theorem is that the vacuum (ground state) expectation value of a product of an \emph{even} number of fermionic operators $A_i$ reduces to a sum of elementary contractions \cite{Molinari17}
\begin{equation}\label{e:wick}
\mel{\psi_i}{A_1 \ldots A_{2n} }{\psi_i} = \sum_{\pi \in \Pi} \mathrm{sgn}(\pi) \expval{A_{i_1}A_{j_1}} \ldots \expval{A_{i_n}A_{j_n}} ,
\end{equation}
where $\Pi \subset S_{2n}$ is the set of permutations $\{ \pi \colon \{ 1,2, \ldots ,2n \} \to \{ i_1,j_1,i_2,j_2, \ldots ,i_n,j_n \} \,\vert\, i_1 < i_2 < \ldots < i_n \text{ and } i_k < j_k \text{ for } k = 1,2,\ldots , n \}$ and $\expval{A_{i_k}A_{j_k}} = \mel{\psi_i}{A_{i_k}A_{j_k}}{\psi_i}$.

Second, we note that the right-hand side of \eqref{e:wick} is equivalent to the Pfaffian $\mathrm{Pf}(M)$ of an anti-symmetric $2n \times 2n$ matrix $M$ with entries
\begin{equation}\label{e:M}
M_{rs} = \expval{A_{r}A_{s}} \text{ for } 1 \leq r < s \leq 2n .
\end{equation}
Since $O_{lm}(t)$ is a product of $2(m-l)=2d$ fermionic operators, we then have that $\mathcal{M}(t)= \lvert \mathrm{Pf}(M) \rvert$. Consequently, for a given time $t\geq 0$, calculating \eqref{e:sop2} is reduced to three steps:
\begin{enumerate}[(i)]
	\item \label{s:elcon} Calculate the four distinct types of elementary contractions $\expval{\phi^{p}_a(t) \phi^{q}_b(t)}$ where $l \leq a \leq b \leq m$ and $p,q \in \{ +,-\}$.
	\item Construct the antisymmetric matrix $M \in \CC^{2d \times 2d}$ with entries
	\begin{equation}\label{e:asymM}
	\begin{split}
	M_{rs} &= \expval{\phi^{p}_a(t) \phi^{q}_b(t)} \text{ such that for } 1 \leq r < s \leq 2d\\
	p &= \begin{cases} +& \text{ for } r \text{ even}, \\ -& \text{ for } r \text{ odd}, \end{cases} \\
	q &= \begin{cases} +& \text{ for } s \text{ even}, \\ -& \text{ for } s \text{ odd}, \end{cases} \\
	a &= l + \floor{r/2} \text{ and } b = l + \floor{s/2} ,
	\end{split}
	\end{equation}
	where $\floor{x}$ is the largest integer smaller than or equal to $x \in \RR$.
	\item Calculate $\lvert \mathrm{Pf}(M) \rvert$, which can be done either by numeric calculation of the Pfaffian,\cite{Wimmer11} or by using the relation $\mathrm{Pf}(M)^2 = \det(M)$. Calculating $\lvert \sqrt{\det(M)} \rvert$ is faster, especially for large matrices.
\end{enumerate}
It remains to determine the elementary contractions of step $($\ref{s:elcon}$)$. We show in Appendix~\ref{app:elcont} how to express these in terms of matrix products conducive to numeric implementation. 

For the numeric results of Sec.~\ref{sec:results} we choose lattice sites $l,m$ in \eqref{e:sop1} symmetrically around the center of the lattice, i.e., $l=(N-d)/2$ and $m=(N+d)/2$. All results shown are for $d=N/2$.

\section{Results and discussion}\label{sec:results}
\begin{figure}[!ht]
	\centering
	\hspace{-.25 cm}
	\includegraphics[width=.45\textwidth]{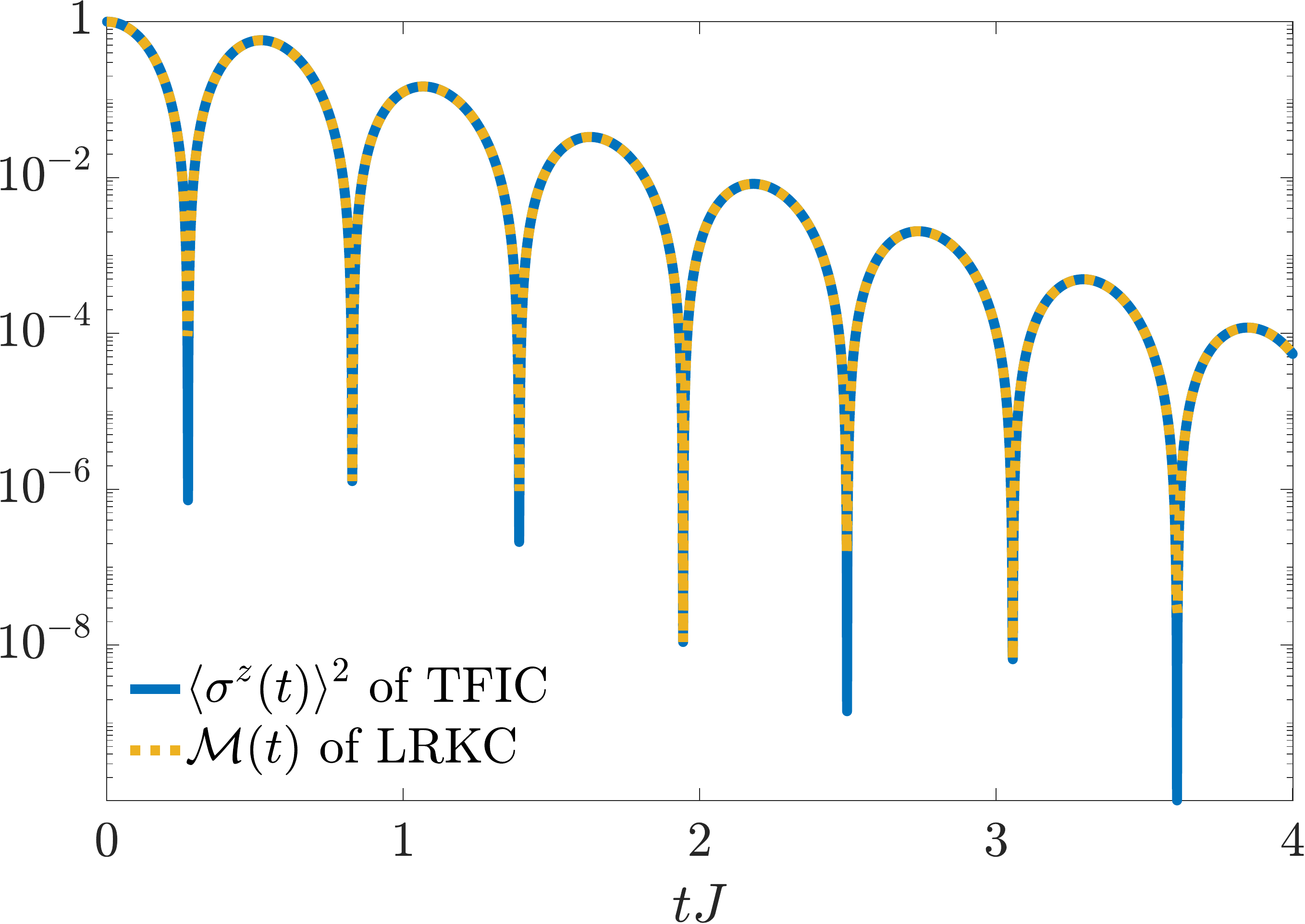}\quad
	\hspace{-.15 cm}
	\caption{(Color online). Time evolution of the squared longitudinal magnetization of the TFIC~\eqref{eq:Ising} and the norm of the string order parameter of its Jordan-Wigner mapping for a quench from $h_i=0$ to $h_f=3$. The result for the magnetization is simulated using TDVP.\cite{Haegeman2011,Haegeman2016,Zauner2018}} 
	\label{fig:FigMot2} 
\end{figure}
Finally, we compare the return rate and the SOP for quenches leading to the three different dynamical phases identified in Sec.~\ref{sec:DPD}, and we discuss the intimate relation between the periodicities of cusps in the return rates and those of the zero crossings of the SOP.
\begin{figure*}[t!]
	\centering
	\hspace{-.25 cm}
	\includegraphics[width=.335\textwidth]{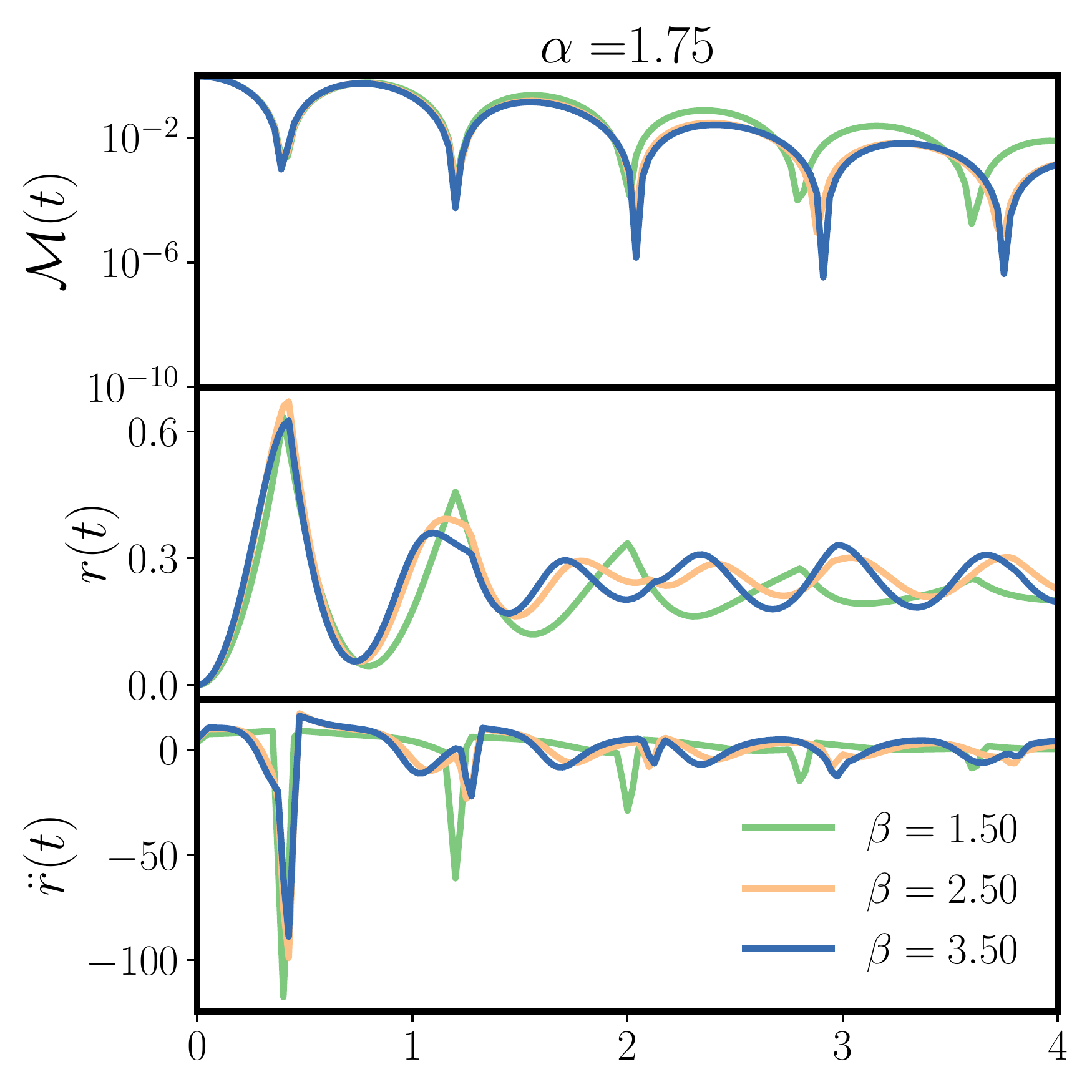}\quad
	\hspace{-.4 cm}
	\includegraphics[width=.335\textwidth]{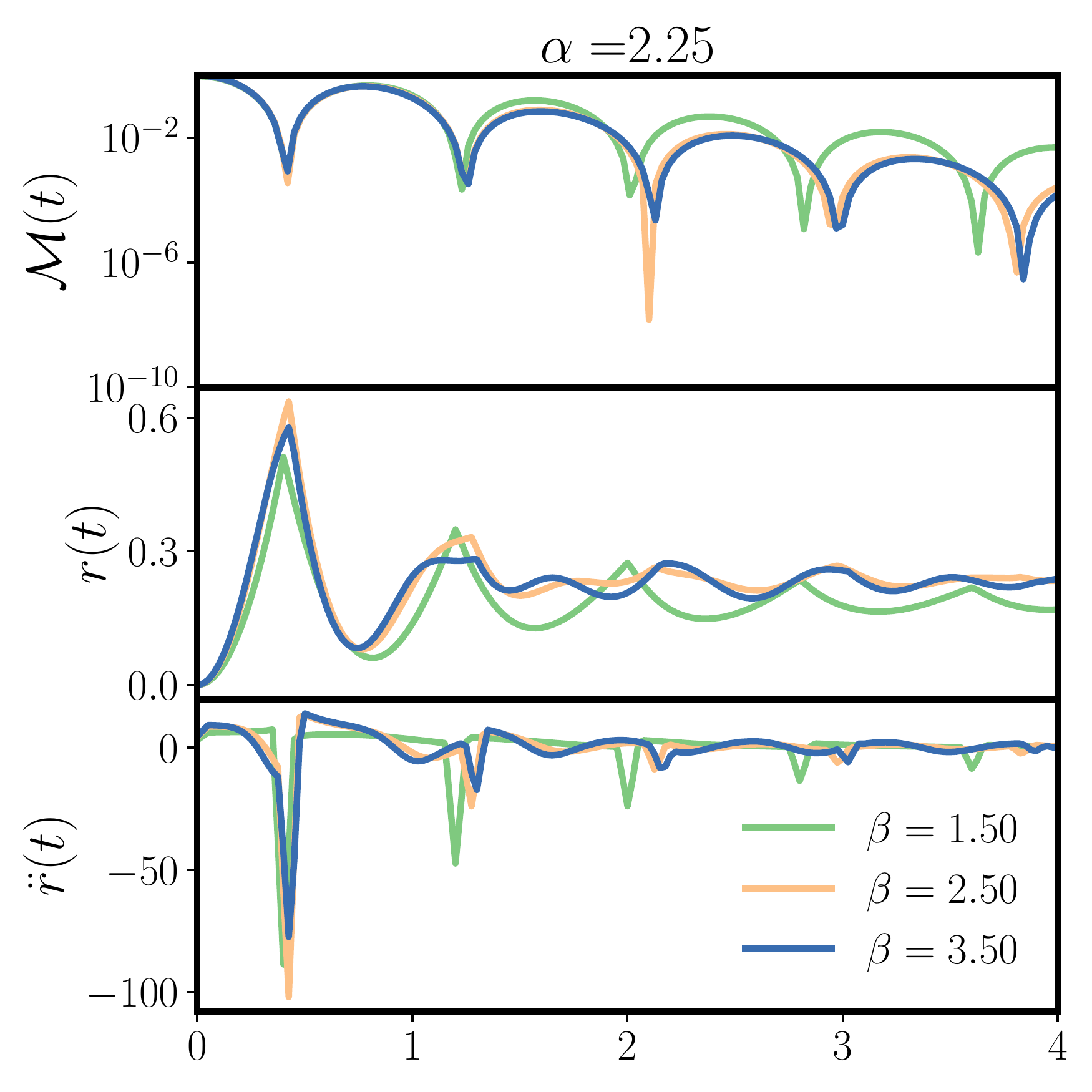}\quad
	\hspace{-.4 cm}
	\includegraphics[width=.335\textwidth]{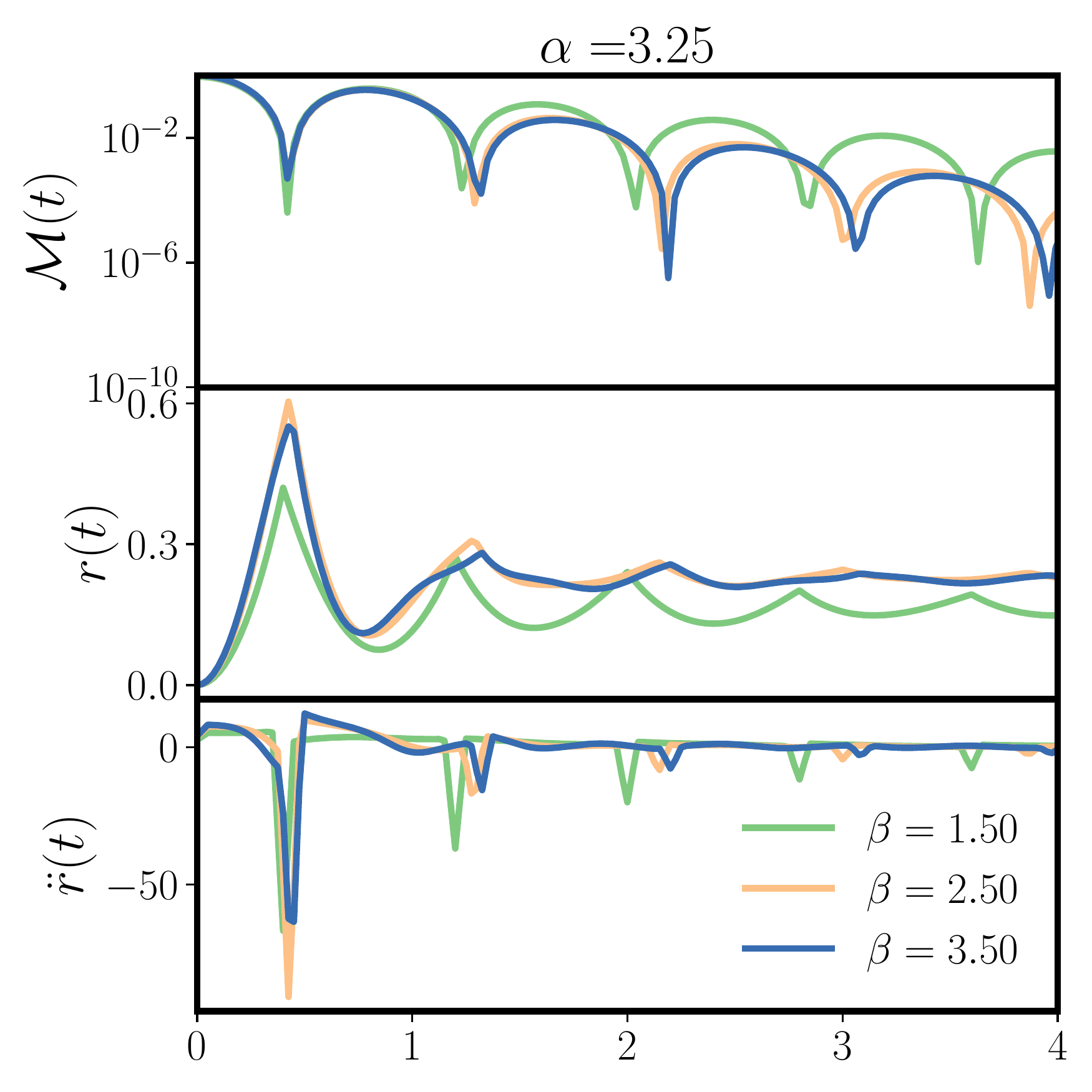}
	\hspace{-.15 cm}
	\caption{(Color online) The time evolution of the return rate $r(t)$ and SOP norm $\mathcal{M}(t)$ for the Kitaev chain with long-range hopping and pairing with decay exponents $\alpha$ and $\beta$, respectively, for three different values of $\alpha\in\{1.75,2.25,3.25\}$, each at three different values of $\beta\in\{1.50,2.50,3.50\}$. These results are for quenches from $h_i=0$ to $h_f>h_c^e$, i.e., in the singly critical dynamical phase. The SOP is calculated for $N=4000$ sites while the return rate $r(t)$ for $N=10000$ sites, where these system sizes are chosen such that the corresponding quantity reaches convergence. The second time-derivative (curvature) of the return rate, $\ddot{r}(t)$, is also provided in order to highlight cusps that may not appear clearly in $r(t)$. As can be seen, the cusps of $r(t)$ and the zeros of $\mathcal{M}(t)$ exhibit the same periodicity.} 
	\label{Fig1} 
\end{figure*}

 It has already been shown in the nearest-neighbor case that for quenches of the chemical potential $h$ across the equilibrium phase boundary $h^{e}_{c}$ the kinks appearing in the return rate correspond to zeros of the string order parameter's norm $\mathcal{M}(t)$.\cite{Budich2016} In Fig.~\ref{fig:FigMot2} we show the time evolution of the square of the TDVP-computed longitudinal magnetization in the TFIC~\eqref{eq:Ising} and that of the norm of the string order parameter in the corresponding Jordan-Wigner mapping, which is the LRKC Hamiltonian~\eqref{eq:lrkc} in the limit of $\alpha,\beta\to\infty$. When $\alpha$ and $\beta$ are finite, we can no longer in general find a spin model that can be mapped onto the corresponding LRKC Hamiltonian~\eqref{eq:lrkc}. Consequently, the LRKC can no longer be associated with a local order parameter such as the longitudinal magnetization. Nevertheless, the fact that the SOP norm and the square of the magnetization are the same observable in the nearest-neighbor limit motivates the study of the SOP~---~which is accessible regardless of the value of $\alpha$ and $\beta$~---~as a viable (nonlocal) order parameter. It further motivates relating its dynamics to that of the return rate, as is done in the case of the local order parameter in spin models.
 
 The picture in the long-range case is similar to its nearest-neighbor counterpart in that large quenches across $h_c^e$ lead to SOP zeros and return rate cusps that exhibit the same periodicity. This is shown in Fig.~\ref{Fig1} where we calculate the dynamical evolution of the system after a quench of the chemical potential to $h=h_{f}=2$ starting from the ground state of the $h=h_i=0$ Hamiltonian for different values of $\alpha$ and $\beta$. The return rate and SOP both display regular singularities in the form of cusps and zero crossings, respectively. As in the nearest-neighbor case,\cite{Budich2016} these singularities occur simultaneously. It is remarkable that the time evolution of the return rate remains very regular (and is similar to the nearest-neighbor case) in the case of $\beta<\alpha$, regardless of the actual $\beta$ value.

\begin{figure*}[!ht]
	\centering
	\hspace{-.25 cm}
	\includegraphics[width=.4\textwidth]{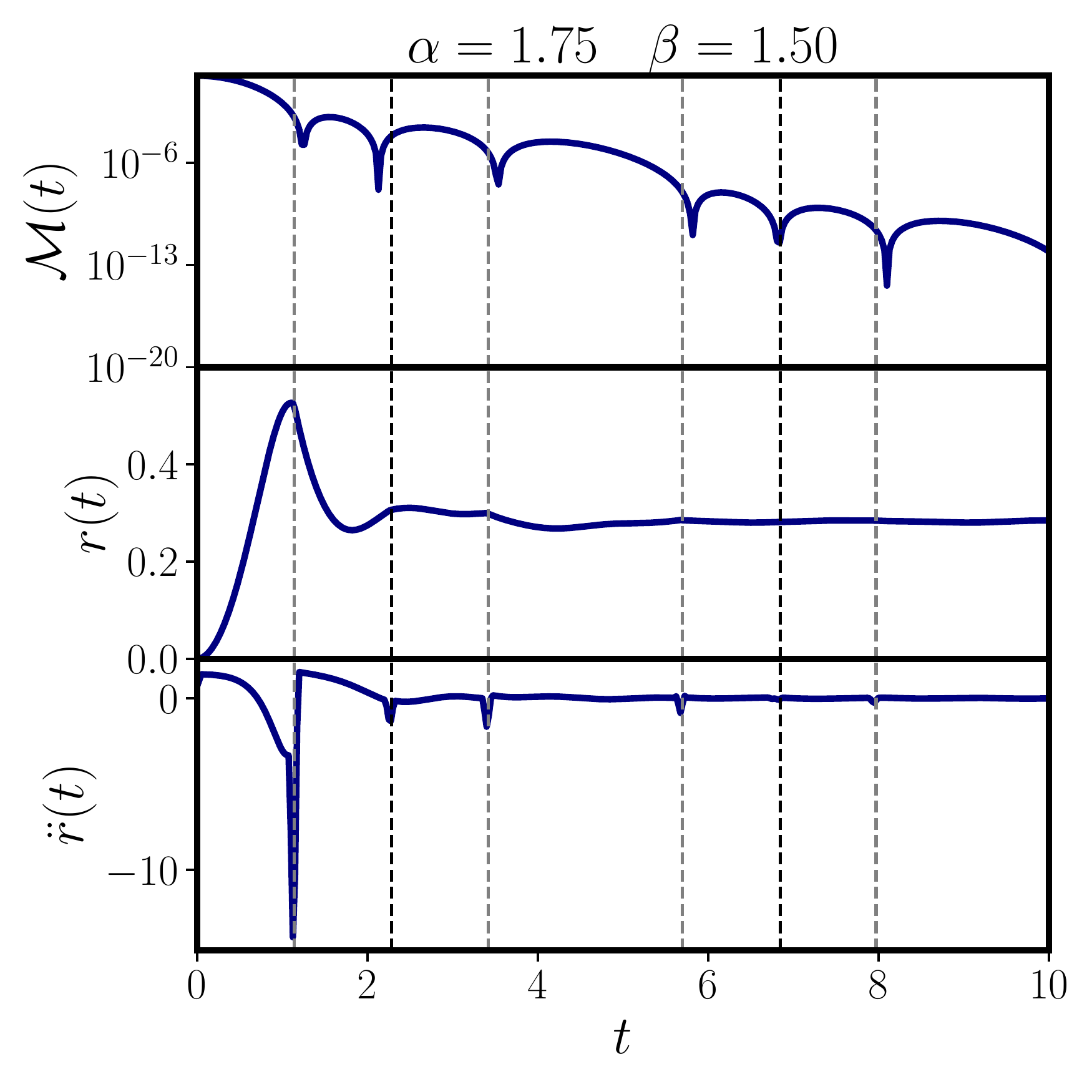}\quad
	\hspace{-.4 cm}
	\includegraphics[width=.4\textwidth]{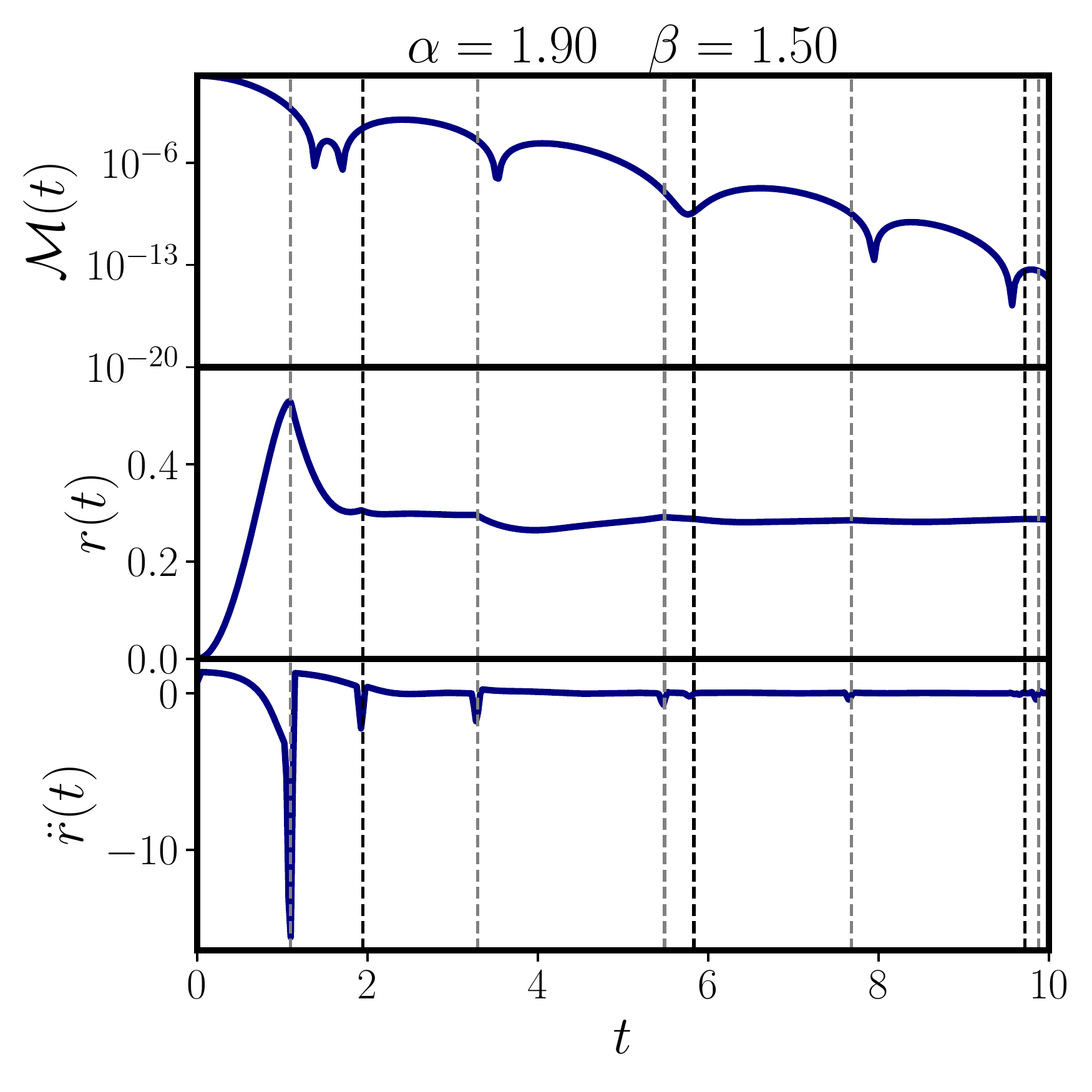}
	\hspace{-.15 cm}
	\caption{(Color online). Return rate and SOP norm for a quench from $h_i=0$ into the the \textit{doubly} critical phase with $h_f=1.075 h_c^d<h_c^e$, for two choices of $(\alpha,\beta)$. In both panels the top quadrant shows the SOP norm and the middle quadrant the return rate. The bottom quadrant shows the curvature $\ddot{r}(t)$ of the return rate in order to identify the critical times, which are further indicated by the vertical dashed lines. The critical times exhibit two distinct periodicities, distinguished by the black and gray dashed lines, respectively, arising from the two distinct critical momenta of this quench. The system sizes were chosen as in Fig.~\ref{Fig1}.} 
	\label{fig:doubly} 
\end{figure*}
\begin{figure*}[!ht]
	\centering
	\hspace{-.25 cm}
	\includegraphics[width=.4\textwidth]{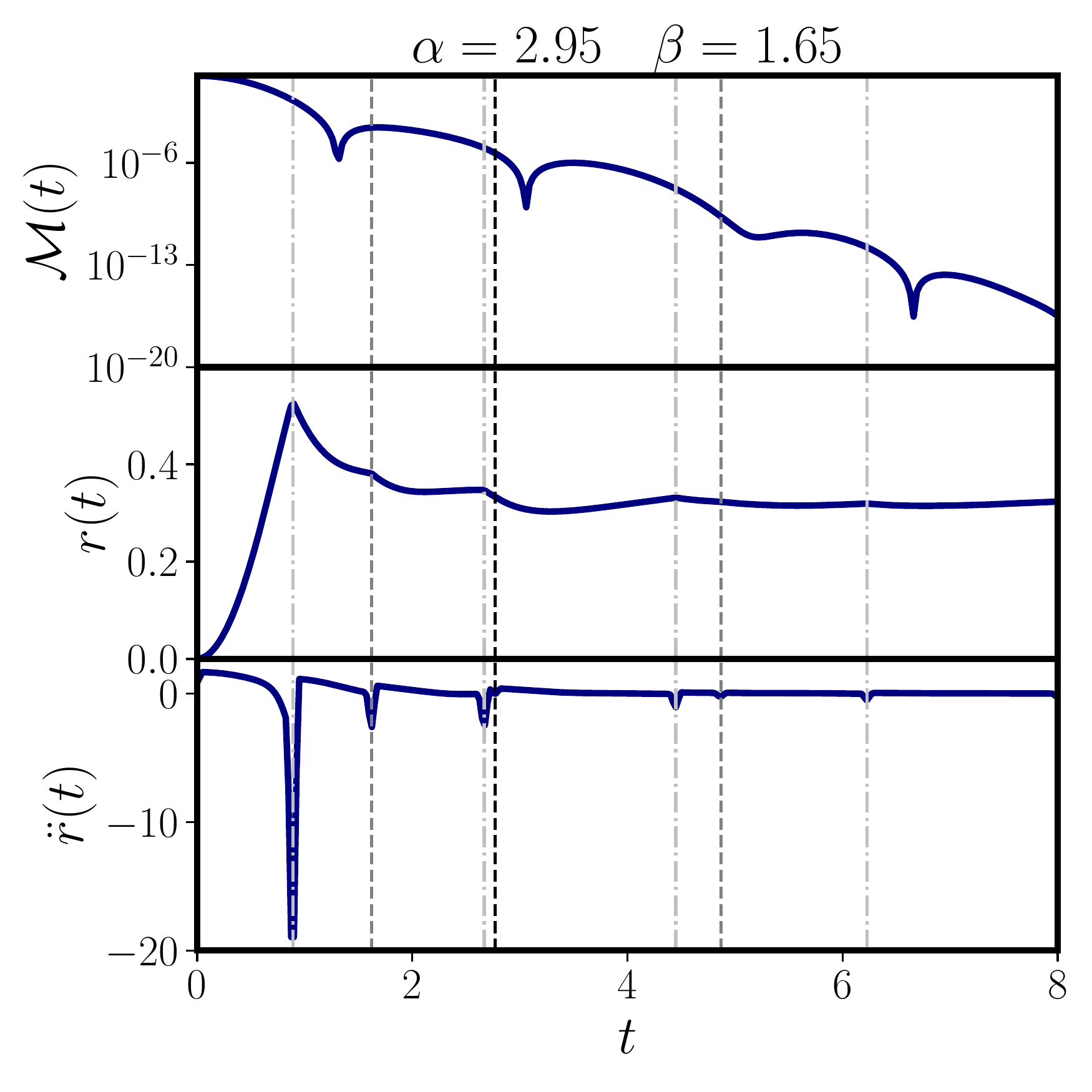}\quad
	\hspace{-.4 cm}
	\includegraphics[width=.4\textwidth]{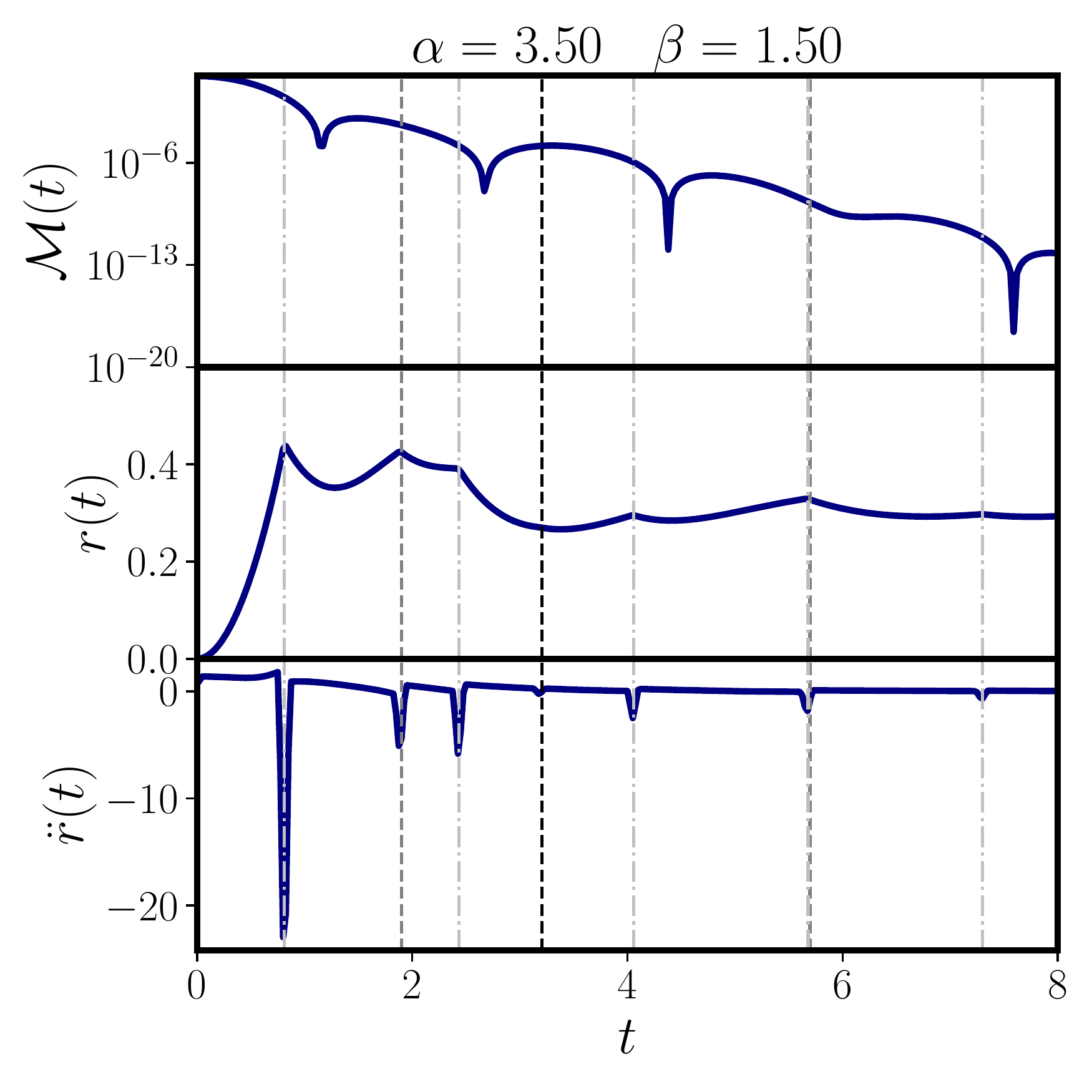}
	\hspace{-.15 cm}
	\caption{(Color online). Return rate and SOP norm for a quench from $h_i=0$ into the the \textit{triply} critical phase with $h_f= 1.05>h_c^e$, for two choices of $(\alpha,\beta)$. The quadrants are arranged as in Fig.~\ref{fig:doubly}. The vertical lines indicate the critical times with three distinct periodicities (black dashed, dark gray dashed, light gray dot-dashed) arising from the three critical momenta of this quench. The system sizes were chosen as in Fig.~\ref{Fig1}.} 
	\label{fig:triply} 
\end{figure*}

Having identified the doubly critical phase boundary $h_{c}^{d}$, we can study the dynamical properties of this phase by quenching the system from $h_{i}=0$ to the final field $h_{f}=1.075h_{c}^{d}<h_c^e$, where two dynamical critical momenta appear. Not surprisingly, for small $\alpha$ and $\beta$, numerical calculations are plagued by considerable finite-size effects. Therefore, we will focus on two particular sets of decay rates $(\alpha,\beta)=(1.75,1.50)$ and $(1.90,1.50)$, both of which are inside the relevant region where the effect of long-range couplings is expected to be more drastic and the doubly critical phase is larger; cf.~Fig.\,\ref{Fig2}. Indeed, as per the analysis of Sec.~\ref{sec:DPD}, in both these cases $\beta<2$ and $\alpha<2\beta-1$, meaning that for quenches just below the equilibrium critical point a doubly critical phase emerges, while for quenches above it, only the singly critical phase can appear; cf.~top right panel of Fig.~\ref{fig:DPD}. The dynamics of the return rate and the SOP for these two parameter sets are shown in Fig.~\ref{fig:doubly}. Even though not as clear cut as in the case of the singly critical phase, we see that the cusps in the return rate and the zero crossings of the SOP nevertheless share similar periodicities. Moreover, the number of cusps over the time interval we show equals the number of zero crossings the SOP makes. Once again, this indicates a nontrivial connection between DQPT and the dynamics of the SOP, much the same way the local order parameter in spin chains has been shown to be intimately connected to regular cusps in the return rate.\cite{Halimeh2017,Zauner2017,Homrighausen2017,Lang2017,Lang2018,Halimeh2018a,Hashizume2018} We note here how both the cusps of the return rate and the zero crossings of the SOP show two periodicities due to these quenches having two emerging critical momenta. These periodicities are indicated by two different fonts of dashed lines, each corresponding to the critical times of a given periodicity.

It is interesting here to relate the doubly critical phase we observe in our simulations to recent experimental findings. In the experiment of Ref.~\onlinecite{Flaeschner2018}, a topological Haldane-like model is implemented using spin-polarized fermions in a driven optical lattice. Initially suppressed with a large energy offset, the tunneling strength between the two sublattices is then suddenly quenched, where near-resonant driving reestablishes tunneling in the final Floquet Hamiltonian. A time-resolved Bloch state tomography is then carried out for the pseudo-spin-$1/2$ in momentum space arising from the sublattice degree of freedom. This system gives rise to the same physics of spin chains in the limit of no interactions. Dynamical vortices~---~which indicate that the time-evolved wavefunction is orthogonal to the initial state, and are therefore representative of cusps in the Loschmidt return rate~---~are directly observed when quenching between two phases of different Chern numbers. However, \textit{accidental} dynamical vortices are also observed for quenches close to the above phase transition, but not crossing it. Indeed, the right panels of Fig.~\ref{fig:DPD} show two forms of $h_f(k)$ described in~\eqref{eq:hf_reduced} that lead to the dynamical phase diagram hosting a doubly critical phase in which nonanalytic behavior in the return rate arises for quenches within the topologically nontrivial phase in the LRKC. This could very well be the underlying phenomenon behind the accidental dynamical vortices in Ref.~\onlinecite{Flaeschner2018}. One possible way of ascertaining this would be to study these accidental dynamical vortices as a function of varying the range of the hopping terms of the Floquet system into which they quench. As detailed through our analytic treatment and numerical results in Sec.~\ref{sec:DPD}, there are certain regimes of the hopping and pairing exponents $(\alpha,\beta)$ where the doubly critical phase is prominent~---~such as in forms (ii) and (iv) in Fig.~\ref{fig:DPD}~---~or outright absent~---~such as in forms (i) and (iii) in Fig.~\ref{fig:DPD}. In particular, as proven in Sec.~\ref{sec:DPD} and illustrated in Fig.~\ref{Fig5}, when $\alpha$ is sufficiently small (i.e., the hopping is sufficiently long-range), the dynamical critical phase will emerge regardless of the value of $\beta$, even when pairing is nearest-neighbor. Therefore, by making $\alpha$ smaller, more of the abovementioned accidental dynamical vortices are expected to appear deeper within the same topological phase and for significantly smaller quench distances. 

However, here we have to be careful and note that there are nontrivial differences between our model and that realized in Ref.~\onlinecite{Flaeschner2018}. Not only is their model two-dimensional, but their quenches also always start in the topologically trivial phase. In contrast, for the case of the (one-dimensional) LRKC that we consider here, quenches within the topologically trivial phase do not seem to give rise to cusps in the return rate (not shown, but extensively studied by us). Nevertheless, this may be different if the LRKC is extended to two dimensions as in the experimental model. In this case, it may be possible that quenches within the topologically trivial phase, and not just its nontrivial counterpart, will also lead to nonanalytic behavior in the return rate due to higher overall connectivity. We leave this question open for possible future investigation.

Finally, we consider quenches that give rise to the triply critical dynamical phase in the LRKC. Unlike the case of the doubly critical phase, and similar to that of the singly critical phase, the triply critical phase occurs for quenches crossing $h_c^e$, albeit it seems to only appear when the quench is right above the equilibrium critical point, $h_f\gtrsim h_c^e$, and also when the model is in the relevant region of long-range pairing $\beta<2$; cf.~Sec.~\ref{sec:triply}. Note that the region of $h_f$ to which one must quench in order to observe the triply critical phase can either be immediately above the equilibrium point up to a small $h_f>1$ (see form (iv) of $h_f(k)$ in bottom right panel of Fig.~\ref{fig:DPD}), or within a small region slightly above, but separated from, $h_c^e$ (see form (iii) of $h_f(k)$ in bottom left panel of Fig.~\ref{fig:DPD}). In Fig.~\ref{fig:triply} three different fonts of dashed lines indicate the critical times in the return rate arising due to three distinct critical momenta for these quenches. Here, the connection with the zero crossings in the SOP is more subtle than in the case of the singly and doubly critical phases. The fact that the triply critical phase appears in a very small regime of $(\alpha,\beta)$ makes the investigation of this subtlety more challenging. This is mainly because with $h_f$ so close (from above) to the equilibrium critical point, most quenches would require impractically large lattice sizes in order to obtain converged dynamics due to critical fluctuations. Nevertheless, it is still not completely surprising here that the connection is not one-to-one as in the singly critical phase. As briefly discussed in Sec.~\ref{sec:sop}, when long-range interactions are present in quantum spin chains, the connection between cusps in the return rate and zeros in the order parameter becomes less straightforward compared to the nearest-neighbor case.\cite{Homrighausen2017,Lang2018} This has been clearly demonstrated in the LMG model, which is the fully connected transverse-field Ising model, where especially at short times the periodicity of the cusps in the return rate does not match that of the zero crossings of the order parameter, in stark contrast to the case of the TFIC. At longer times, these two periodicities nevertheless do become equal in the LMG model. In the case of LRKC, the SOP decreases exponentially fast, making it much harder to tell whether this may also be the case here.

\section{Conclusion}\label{sec:conc}
We have investigated the effect of having both long-range hopping ($\propto1/r^\alpha$) and long-range pairing ($\propto1/r^\beta$) on the dynamics of free-fermionic models in the wake of a quench. In particular, we have thoroughly constructed the rich dynamical phase diagram of the long-range Kitaev chain using scaling arguments and extensive numerical simulations. We find three main phases: a singly critical phase known from seminal works in which the cusps of the return rate correspond to a single critical momentum mode; the doubly critical phase in which cusps in the return rate correspond to two distinct critical momenta; and the triply critical phase where cusps arise due to three distinct critical momenta. We identified regimes of $(\alpha,\beta)$ associated with each of these phases, where the succinct summary would be that small $\alpha$ guarantees the emergence of the doubly critical phase, while small enough $\beta$ is necessary for the triply critical phase to exist. Both the singly and the triply critical phases can occur for quenches crossing the equilibrium critical point, whereas in contrast the doubly critical phase appears for quenches within the topologically nontrivial phase.\begin{figure*}[!ht]
	\centering
	\hspace{-.25 cm}
	\includegraphics[width=.45\textwidth]{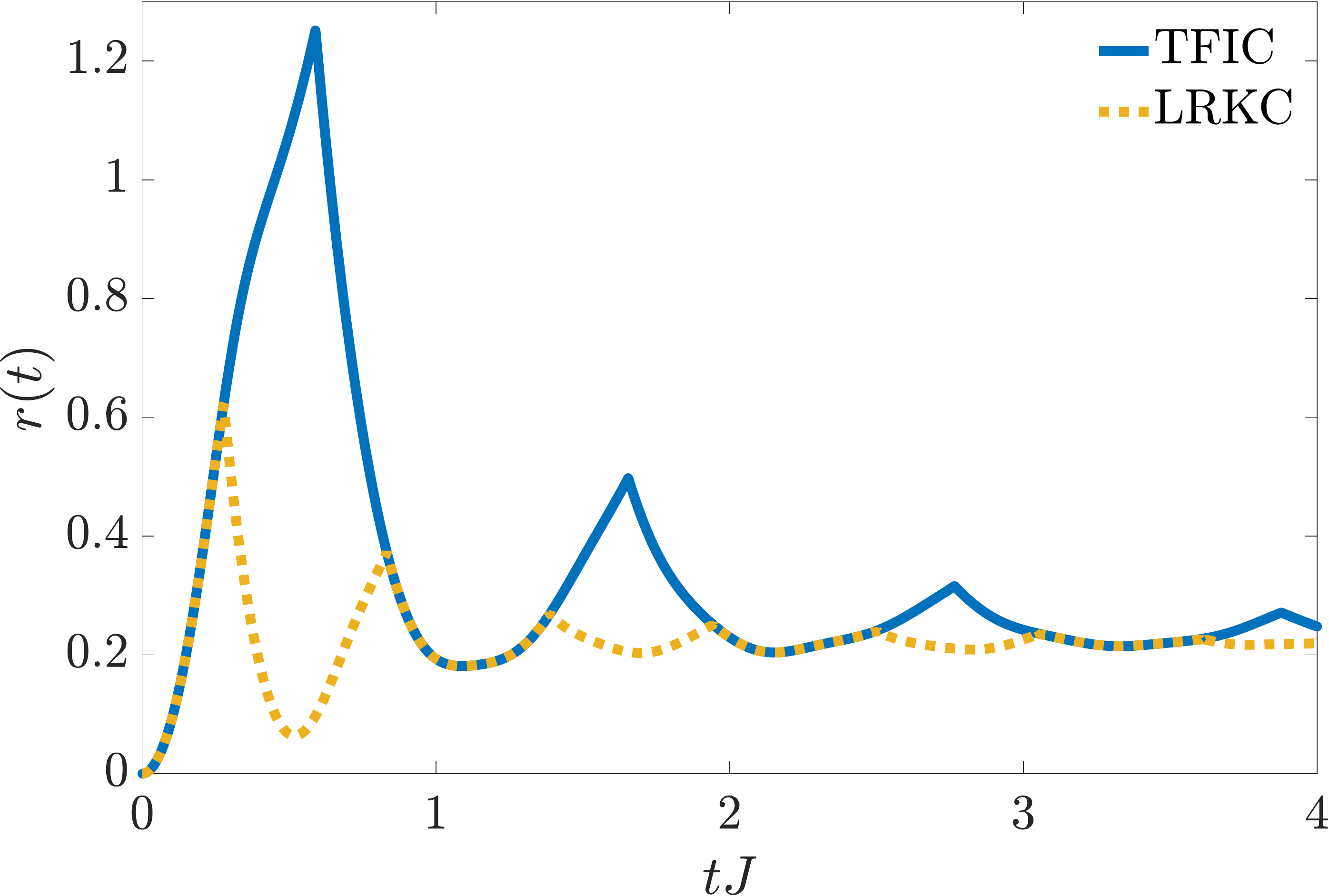}\quad
	\hspace{-.4 cm}
	\includegraphics[width=.45\textwidth]{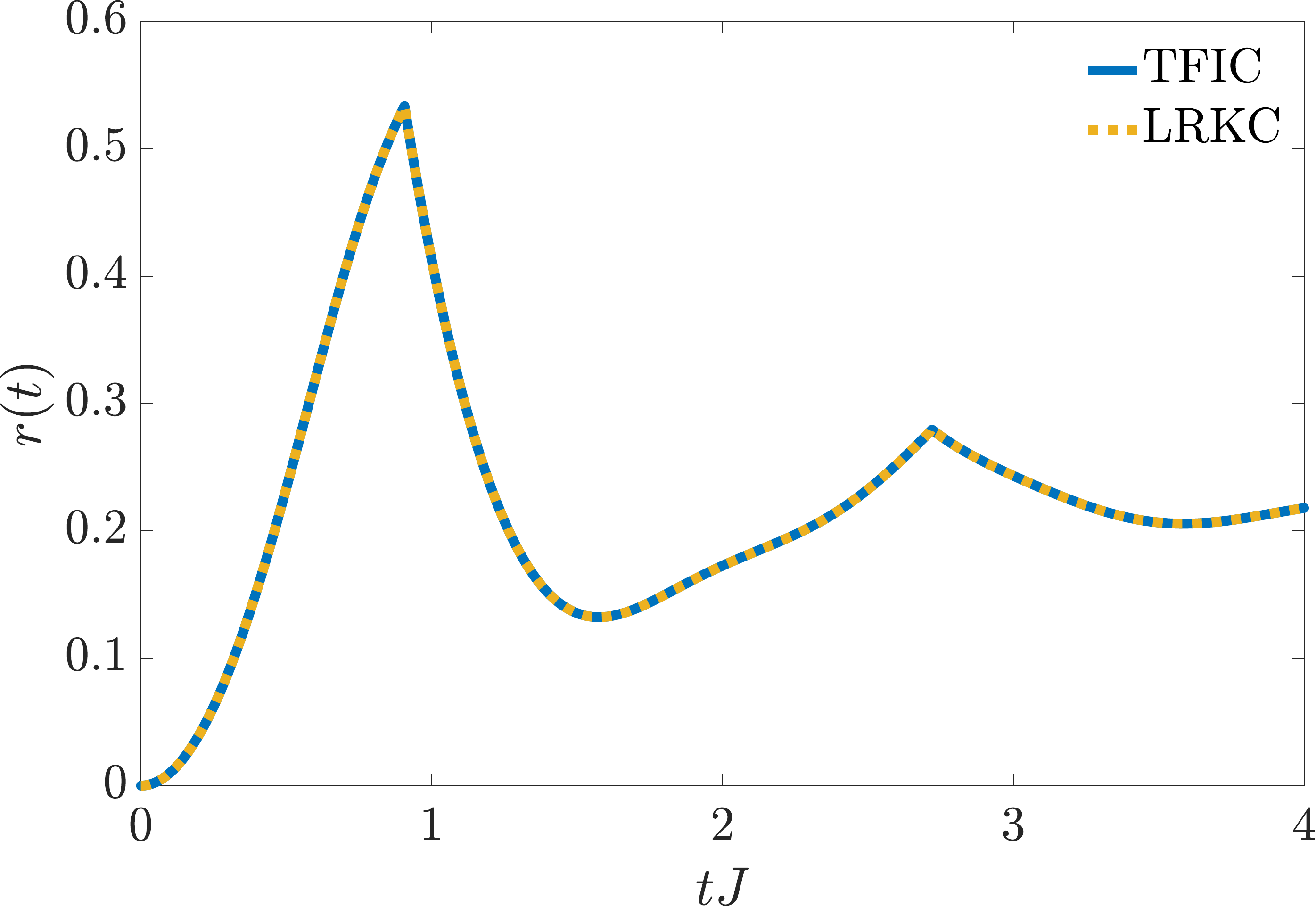}
	\hspace{-.15 cm}
	\caption{(Color online). Comparison between the return rate of the TFIC~\eqref{eq:Ising} and that of the LRKC~\eqref{eq:lrkc} in the limit $\alpha,\beta\to\infty$ for a quench from $h_i=0$ to $h_f=3$ (left panel) and a quench from $h_i\to\infty$ to $h_f=0.5$ (right panel). The return rates of the TFIC are computed using TDVP.} 
	\label{fig:comp} 
\end{figure*}

Furthermore, we have related the cusps in the Loschmidt return rate to the zero crossings of the string order parameter. Our results indicate that there is a close connection between the cusps and zero crossings, especially in the case of singly and doubly critical dynamical phases. For large quenches from the topologically nontrivial phase crossing the equilibrium critical point, a singly critical phase arises regardless of the values of $\alpha$ and $\beta$. This extends what is already known on DQPT as a probe of the underlying equilibrium physics, but also shows that the connection between DQPT and equilibrium physics is not one-to-one, especially in the case of the doubly critical phase as it arises for quenches within the topologically nontrivial phase, i.e., without having to cross any equilibrium critical point.

Our work shows experimental promise as it may explain the so-called \textit{accidental} dynamical vortices from time-resolved Bloch tomography measurements in Ref.~\onlinecite{Flaeschner2018} that appear for quenches close to, but not crossing, the topological phase transition. This parallels our observation of cusps in the return rate for quenches within the same topological phase, giving rise to a doubly critical dynamical phase possessing two distinct periodicities in the return rate cusps. Our analysis suggests that by tuning the hopping range these accidental vortices can be better understood in that for longer hopping ranges more vortices will appear deeper within the same topological phase and for smaller quenches. 

Finally, it is worth mentioning that our results suggest that the SOP can offer an indirect way of measuring the nonanalyticities of the Loschmidt return rate. In the work of Refs.~\onlinecite{Endres2011,Hilker17}, a quantum gas microscope has been used to measure the string order, respectively, in one-dimensional bosonic Mott insulators and the Fermi-Hubbard chain. Given the close connection evident in our simulations between the return rate cusps and the SOP zeros, this would be a viable and much more practical way of observing dynamical critical behavior than actually measuring the return rate itself, which becomes exponentially harder with system size.

\section*{Acknowledgments}
J.C.H.~is grateful to Jan Carl Budich, Markus Heyl, and Sheldon T.~Joyner for stimulating discussions and valuable comments on various aspects of this project. J.C.H.~is grateful to Christof Weitenberg for fruitful discussions, and for valuable comments on and a thorough reading of our manuscript. J.C.H.~and P.U.~acknowledge support by the ERC Starting Grant StrEnQTh. This work has been supported by the Deutsche Forschungsgemeinschaft (DFG) via Collaborative Research Centre SFB 1225 (ISOQUANT) and under Germany's Excellence Strategy EXC2181/1-390900948 (Heidelberg STRUCTURES Excellence
Cluster).

\appendix

\section{Nearest-neighbor LRKC and TFIC: return rate}\label{app:compare}
In Fig.~\ref{fig:comp} we compare the return rates of the TFIC~\eqref{eq:Ising} and the LRKC~\eqref{eq:lrkc} in the limit $\alpha,\beta\to\infty$. These two models are identical via an exact Jordan-Wigner mapping.

We see that there is a qualitative difference between the return rates in the case of a quench from the ordered phase ($h_i<h_c^d$), with the periodicity of cusps in the LRKC being half that of the TFIC. This is due to the fact that in the fermionic system the ground state is always nondegenerate, while in the case of the TFIC, the ground state in the ferromagnetic phase is always doubly degenerate. The return rates can be brought to quantitative and qualitative agreement by updating the return rate definition to

\begin{align}
r(t)=-\lim_{N\to\infty}\frac{1}{N}\ln\sum_{n=0}^1|\langle\psi_{i,n}|\mathrm{e}^{-\mathrm{i}H_ft}|\psi_{i,n}\rangle|^2,
\end{align}
where $\{\ket{\psi_{i,0}},\ket{\psi_{i,1}}\}$ are the two degenerate ground states at $h_i<h_c^e$.

On the other hand, the return rates are in full qualitative and quantitative agreement in the case of quenches from the paramagnetic phase. This is because in the latter phase the TFIC has a nondegenerate ground state.

\section{Elementary contractions for the numeric calculation of string order parameters}\label{app:elcont}
Here we demonstrate how to calculate the elementary contractions $\expval{\phi^{p}_a(t) \phi^{q}_b(t)}$, required in step $($\ref{s:elcon}$)$ of Sec.~\ref{sec:sop}.

We will require two linear canonical transformations which respectively diagonalise the pre- and post-quench Hamiltonians as
\begin{equation}\label{e:gammaeta}
H_i = \sum_{n=1}^N \lambda_{n}^i \alpha_n^\dagger \alpha_n \text{ and }	H_f = \sum_{n=1}^N \lambda_{n}^f \beta_n^\dagger \beta_n,
\end{equation}
The motivation for this is that whilst the post-quench evolution of the fermionic operators $\phi_j^{\pm}(t)$ is easily solved in the post-quench $\beta$-basis, we would like to make use of the fact that in the pre-quench $\alpha$-basis terms such as $\alpha_{n} \ket{\psi_i}$ vanish trivially since $\ket{\psi_i} = \ket{0}_\alpha$.

We define the relevant \emph{real} transformation matrices as
\begin{align}\label{e:lintrans}
\begin{pmatrix}
\vec{\beta} \\ \vec{\beta^\dagger}
\end{pmatrix}
&= 
\begin{pmatrix} G_f & F_f \\ F_f & G_f	\end{pmatrix}
\begin{pmatrix}
\vec{c} \\ \vec{c^\dagger}
\end{pmatrix},\\
\begin{pmatrix}
\vec{\alpha} \\ \vec{\alpha^\dagger}
\end{pmatrix}
&= 
\begin{pmatrix} G_i & F_i \\ F_i & G_i	\end{pmatrix}
\begin{pmatrix}
\vec{c} \\ \vec{c^\dagger}
\end{pmatrix},
\end{align}
where $\vec{c} = \begin{pmatrix} c_1 & c_2 & \ldots & c_N \end{pmatrix}^\intercal$, $\vec{c^\dagger} = \begin{pmatrix} c^\dagger_1 & c^\dagger_2 & \ldots & c^\dagger_N \end{pmatrix}^\intercal$ are column vectors and, as before, the subscripts $i,f$ respectively refer to the pre- and post-quench values of the chemical potential $h$. Note that since the above transformations are canonical, i.e.\ preserve the fermionic anticommutation relations, matrices $G,F$ satisfy identities
\begin{equation}\label{e:canon}
G^\intercal G + F^\intercal F = \mathds{1}_{N\times N} \text{ and } G^\intercal F + F^\intercal G = 0_{N\times N}.
\end{equation}

One can then obtain the transformation matrices \eqref{e:lintrans} as follows:
First write the long-range Kitaev Hamiltonian \eqref{eq:lrkc} as
\begin{equation}\label{e:HAB}
\begin{split}
H_f =& \sum_{m,n =1}^{N} c_m^\dagger (A_f)_{mn} c_n \\
&+ \frac{1}{2}\left( c_m^\dagger (B_f)_{mn} c_n^\dagger - c_m (B_f)_{mn} c_n \right) ,
\end{split}
\end{equation}
where

\begin{align}
(A_f)_{mn}&=(A_f)_{nm} \in \RR^{N \times N},\\
(B_f)_{mn}&=-(B_f)_{nm} \in \RR^{N \times N}.
\end{align} The exact epressions for these matrices are given below in \eqref{e:AB}.
As shown in Ref.~\onlinecite{LiebSchultzMattis1961} we can then express
\begin{equation}
G_f = (\Phi_f + \Psi_f)/2 \text{ and } F_f = (\Phi_f - \Psi_f)/2 ,
\end{equation}
where the $n$th columns of the real, orthogonal matrices $\Phi_f,\Psi_f$ are obtained by solving the eigensystem
\begin{equation}
(\Phi_f)_n(A_f-B_f)(A_f+B_f) = (\lambda_{n}^f)^2 (\Phi_f)_n ,
\end{equation}
and by making use of the relation
\begin{equation}\label{e:phipsi}
(\Phi_f)_n(A_f-B_f) = \lambda_n^f (\Psi_f)_n .
\end{equation}
The advantage of this approach is that we obtain the transformation matrices of \eqref{e:lintrans} by solving the eigensystem of an $N \times N$ symmetric matrix $(A_f-B_f)(A_f+B_f)$, instead of the $2N \times 2N$ Hamiltonian $H_f$. To obtain $G_i,F_i$ we repeat \eqref{e:HAB}--\eqref{e:phipsi} with $h_f$ replaced by $ h_i$.

Having obtained all transformation matrices we now solve the time evolution of $\phi_b^q(t) \ket{\psi_i}$. Using \eqref{e:canon} to invert \eqref{e:lintrans}, we may write
\begin{equation}\label{e:elcon1}
\begin{split}
\phi_b^q(t) =& \left[ \vec{c^\dagger(t)} + q \vec{c(t)} \right]_b \\
=& \Bigl\{ (F_f)^\intercal e^{-\mathrm{i}\Lambda t} \vec{\beta} + (G_f)^\intercal  e^{\mathrm{i}\Lambda t}  \vec{\beta^\dagger} \\
&+ q\left[ (G_f)^\intercal  e^{-\mathrm{i}\Lambda t} \vec{\beta} + (F_f)^\intercal e^{\mathrm{i}\Lambda t}  \vec{\beta^\dagger} \right] \Bigr\}_b ,
\end{split}
\end{equation}
where we have used \eqref{e:gammaeta} to write $ \vec{\beta(t)} = e^{-\mathrm{i}\Lambda t} \vec{\beta}$ with ${\Lambda = \mathrm{diag}(\lambda_1^f, \ldots, \lambda_N^f)}$. Having solved the time-evolution in the post-quench $\beta$-basis, we use \eqref{e:lintrans} and \eqref{e:canon} to transform directly into the pre-quench $\alpha$-basis as
\begin{equation}\label{e:lintrans2}
\begin{split}
\begin{pmatrix}
\vec{\beta} \\ \vec{\beta^\dagger}
\end{pmatrix}
=&
\begin{pmatrix} T_{1} & T_{2} \\ T_{2} & T_{1}	\end{pmatrix}
\begin{pmatrix}
\vec{\alpha} \\ \vec{\alpha^\dagger}
\end{pmatrix} \text{ with}\\
T_1 =& G_f (G_i)^\intercal + F_f (F_i)^\intercal, \\
T_2 =& G_f (F_i)^\intercal + F_f (G_i)^\intercal .
\end{split}
\end{equation}
This allows us to make use of $\alpha_n \ket{\psi_i}=0$ after substituting \eqref{e:lintrans2} into \eqref{e:elcon1}. We then obtain
\begin{equation}\label{e:elcon2}
\begin{split}
\phi_b^q(t) \ket{\psi_i} =& \left[ (\Theta_f^q)^\intercal \left( q e^{-\mathrm{i}\Lambda t} T_2 + e^{\mathrm{i}\Lambda t} T_1 \right) \vec{\alpha^\dagger} \right]_b \ket{\psi_i},
\end{split}
\end{equation}
where
\begin{equation}
\Theta_f^q = G_f + q F_f = \begin{cases}
\Phi_f, & q=+, \\
\Psi_f, & q=-.
\end{cases}
\end{equation}
Repeating \eqref{e:elcon1}--\eqref{e:elcon2} for $\bra{\psi_i} \phi_a^p(t)$ then yields
\begin{equation}\label{e:elconfinal}
\begin{split}
\mel{\psi_i}{ \phi_a^p(t) \phi_b^q(t)}{\psi_i} =& \Bigl[ (\Theta_f^p)^\intercal \left( p e^{-\mathrm{i}\Lambda t} T_1 + e^{\mathrm{i}\Lambda t} T_2 \right) \\
\times& \left( q (T_2)^\intercal e^{-\mathrm{i}\Lambda t} + (T_1)^\intercal e^{\mathrm{i}\Lambda t} \right) \Theta_f^q  \Bigr]_{ab} ,
\end{split}
\end{equation}
where we have used that for any two $N\times N$ matrices $M_1,M_2$
\begin{equation}
\Bigl\langle \psi_i \Bigr\vert \Bigl[ M_1 \vec{\alpha}  \Bigr]_a \Bigl[ M_2 \vec{\alpha^\dagger}  \Bigr]_b \Bigl\vert \psi_i \Bigr\rangle = \left[ M_1 (M_2)^\intercal \right]_{ab} .
\end{equation}
The useful result here is~\eqref{e:elconfinal}, which expresses the elementary contractions required to populate matrix \eqref{e:asymM} as the elements of a matrix product which is in turn defined purely in terms of the eigensystem of $(A_f-B_f)(A_f+B_f)$ and of $(A_i-B_i)(A_i+B_i)$, which can be easily obtained with, for instance, \textit{Mathematica}.

We conclude by giving the explicit forms for matrices $A$ and $B$ of \eqref{e:HAB}:
\begin{equation}\label{e:AB}
\begin{split}
A_{mn} =& 2h \delta_{m,n} - \sum_{r=1}^{N/2-1} \frac{r^{-\alpha}}{\mathcal{N}_{\alpha}} \bigl[ (-1)^{\Theta(m-N-1+r)} \delta_{m, (n-r)} \\
&+ (-1)^{1-\Theta(m-r-1)} \delta_{m, (n+r)}  \bigr], \\
B_{mn} = & -\sum_{r=1}^{N/2-1}  \frac{r^{-\beta}}{\mathcal{N}_{\beta}} \bigl[ (-1)^{\Theta(m-N-1+r)} \delta_{m, (n-r)} \\
&- (-1)^{1-\Theta(m-r-1)} \delta_{m, (n+r)}  \bigr],
\end{split}
\end{equation}
where $\Theta(x)$ is the Heaviside function and the subscripts of all Kroenecker delta functions $\delta$ are to be read as modulo $N$, i.e.,\ $\delta_{m,(n-r)} = \delta_{m (\mathrm{mod}\, N), (n-r)(\mathrm{mod}\, N)}$ when using circular boundaries. The prefactors expressed in terms of $\Theta(x)$ impose the correct phases under anti-periodic boundary conditions and should be set to $1$ when imposing periodic boundaries instead.
\\\newline

\bibliography{LRKC_biblio}

\begin{thebibliography}{90}
\bibitem[{not()}]{note}
\bibinfo{howpublished}{In fact, the method presented here is applicable to any
  fermionic Hamiltonian that is quadratic in the annihilation and creation
  operators.}
\end{thebibliography}
\end{document}